\documentclass[usenatbib]{mn2e} 
\usepackage{graphics} 
\usepackage{epsfig}
\usepackage{color} 
\usepackage{xspace}
\usepackage{times}

\def\ni{\noindent}                                    

\newcommand{\Ha}{\ifmmode {\rm H}\alpha \else H$\alpha$\fi\xspace}
\newcommand{\Hb}{\ifmmode {\rm H}\beta \else H$\beta$\fi\xspace}
\newcommand{\Hba}{\ifmmode {\rm H}\beta^{\prime} \else H$\beta^{\prime}$\fi\xspace}
\newcommand{\Hg}{\ifmmode {\rm H}\gamma \else H$\gamma$\fi\xspace}
\newcommand{\Hd}{\ifmmode {\rm H}\delta \else H$\delta$\fi\xspace}
\newcommand{\hi}{\ifmmode \rm{H}\,\textsc{i} \else H\,{\sc i}\fi\xspace}

\newcommand{\Hii}{\ifmmode \rm{H}\,\textsc{ii} \else H\,{\sc ii}\fi}

\newcommand{\nii}{\ifmmode [\rm{N}\,\textsc{ii}] \else [N\,{\sc ii}]\fi\xspace}

\newcommand{\oi}{\ifmmode [\rm{O}\,\textsc{i}] \else [O\,{\sc i}]\fi\xspace}

\newcommand{\neiii}{\ifmmode [\rm{Ne}\,\textsc{iii}] \else [Ne\,{\sc iii}]\fi}

\newcommand{\hei}{\ifmmode \rm{He}\,\textsc{i} \else He\,{\sc i}\fi\xspace}
\newcommand{\heii}{\ifmmode \rm{He}\,\textsc{ii} \else He\,{\sc ii}\fi\xspace}
\newcommand{\oii}{\ifmmode [\rm{O}\,\textsc{ii}] \else [O\,{\sc ii}]\fi\xspace}

\newcommand{\oiii}{\ifmmode [\rm{O}\,\textsc{iii}] \else [O\,{\sc iii}]\fi\xspace}

\newcommand{\sii}{\ifmmode [\rm{S}\,\textsc{ii}] \else [S\,{\sc ii}]\fi\xspace}
\newcommand{\siii}{\ifmmode [\rm{S}\,\textsc{iii}] \else [S\,{\sc iii}]\fi}

\def\aj{AJ}
\def\apj{ApJ}
\def\apjl{ApJ}
\def\apjs{ApJS}
\def\aap{A\&A}
\def\aapr{A\&A~Rev.}
\def\aaps{A\&AS}
\def\mnras{MNRAS}

\def\ATT{\color{black}}        

\title[The forgotten population of weak line galaxies in the
SDSS]{Alternative diagnostic diagrams and the ``forgotten'' population
of weak line galaxies in the SDSS}

\author[Cid Fernandes et al]
         {R. Cid Fernandes$^{1}$\thanks{E-mail: cid@astro.ufsc.br},
	  G. Stasi\'nska $^{2}$,
	  M. S. Schlickmann$^{1}$,
 	  A. Mateus$^{1}$,  
	  N. Vale Asari$^{1,2}$,
	 \newauthor
	 W. Schoenell$^{1}$,
	 L. Sodr\'e Jr.$^{3}$
	 (the SEAGal collaboration)\thanks{Semi-Empirical Analysis of Galaxies}\\
	 $^{1}$Departamento de F\'{\i}sica - CFM - Universidade Federal de Santa Catarina,
	 Florian\'opolis, SC, Brazil\\
	 $^{2}$LUTH, Observatoire de Paris, CNRS, Universit\'e Paris Diderot; Place Jules Janssen 92190 Meudon, France\\
	 $^{3}$Instituto de Astronomia, Geof\'{\i}sica e Ci\^encias
         Atmosf\'ericas, Universidade de S\~ao Paulo, S\~ao Paulo, SP,
         Brazil
	 }

\begin{document}

\maketitle

\begin{abstract} 
A numerous population of weak line galaxies (WLGs) is often left out
of statistical studies on emission line galaxies (ELGs) due to the
absence of an adequate classification scheme, since classical
diagnostic diagrams, like \oiii/\Hb\ vs.\ \nii/\Ha (the BPT diagram),
require the measurement of at least 4 emission lines.  This paper aims
to remedy this situation by transposing the usual divisory lines
between Star Forming (SF) and Active Galactic Nuclei (AGN) hosts, and
between Seyferts and LINERs to diagrams that are more economical in
terms of line quality requirements. By doing this, we rescue from the
classification limbo a substantial number of sources and modify the
global census of ELGs. More specifically: (1) We use the Sloan Digital
Sky Survey DR7 to constitute a suitable sample of 280 thousand ELGs,
1/3 of which are WLGs.  (2) Galaxies with strong emission lines are
classified using the widely applied criteria of
\citet{Kewley_etal_2001}, \citet{Kauffmann_etal_2003c},
\citet{Stasinska_etal_2006} and \citet{Kewley_etal_2006}.  (3) We
transpose these classification schemes to alternative diagrams keeping
\nii/\Ha as a horizontal axis, but replacing \Hb\ by a stronger line
(\Ha\ or \oii), or substituting the ionization-level sensitive
\oiii/\Hb ratio with the equivalent width of \Ha\ ($W_{\Ha}$).
Optimized equations for the transposed divisory lines are provided.
(4) We show that nothing significant is lost in the translation, but
that the new diagrams allow one to classify up to 50\% more ELGs.  (5)
Introducing WLGs in the census of galaxies in the local Universe
increases the proportion of metal-rich SF galaxies and especially
LINERs.

In the course of this analysis, we were led to make the following
points: (a) The \citet{Kewley_etal_2001} BPT-line for galaxy
classification is generally ill-used.  (b) Replacing \oiii/\Hb by
$W_{\Ha}$ in the classification introduces a change in the philosophy
of the distinction between LINERs and Seyferts, but not in its
results. Because the $W_{\Ha}$ vs.\ \nii/\Ha diagram can be applied to
the largest sample of ELGs without loss of discriminating power
between Seyferts and LINERs, we recommend its use in further studies.
(c) The dichotomy between Seyferts and LINERs is washed out by WLGs in
the BPT plane, but it subsists in other diagnostic diagrams. This
suggests that the right wing in the BPT diagram is indeed populated by
at least two classes, tentatively identified with bona fide AGN and
``retired'' galaxies that have stopped forming stars and are ionized
by their old stellar populations.
\end{abstract}



\section{Introduction}
\label{sec:Introduction}

Galaxies with emission lines can reveal more of their secrets than
those without.  The strength and pattern of emission lines convey
information on the power and nature of the ionizing source, the
geometry, physical conditions and chemical composition of the gas, as
well as on the dust content of emitting regions. Even when these
properties cannot be determined unambiguously, emission line data
allow one to assign galaxies to physically motivated classes, like
Star-Forming (SF) or hosts of Active Galactic Nuclei (AGN), and to
split them into finer categories such as high (Seyfert) and low
(LINER) ionization sub-types.

The classification of emission-line galaxies is usually done through
emission line ratio diagnostic diagrams.
\citet*{Baldwin_Phillips_Terlevich_1981} were the first to propose
such a scheme. Their \oiii/\Hb versus \nii/\Ha diagram\footnote{We
denote \oii$\lambda\lambda$3726+3729, \oiii$\lambda$5007,
\oi$\lambda$6300, \nii$\lambda$6584 and \sii$\lambda\lambda$6716+6731
by simply \oii, \oiii, \oi, \nii and \sii, respectively.} (the `BPT
diagram'), in particular, became the benchmark for emission line
classification.  In their original plot, HII regions, planetary
nebulae, Seyfert nuclei and LINERs occupied well isolated regions,
demonstrating the diagnostic power of this combination or flux ratios.
As data accumulated through the 80's and 90's
(\citealp{Veilleux_and_Osterborck_1987,Ho_Filippenko_Sargent_1997c,
Veron-Cetty_Veron_2000}), a more continuous distribution gradually
emerged, culminating with the Sloan Digital Sky Survey (SDSS;
\citealp{York_etal_2000}), which allowed the mapping of the BPT plane
with over $10^5$ data points \citep{Kauffmann_etal_2003c}, revealing
the now familiar seagull-like shape, with two well defined wings.

The left wing arises due to a strong coupling between the O/H and N/O
abundance ratios, the ionizing radiation field and the ionization
parameter in SF galaxies \citep*{McCall_Rybski_Shields_1985,
Dopita_Evans_1986}. Empirical and model-based frontiers have been
drawn in the \oiii/\Hb-\nii/\Ha space to delineate the SF territory
from the rest (\citealp{Kewley_etal_2001, Kauffmann_etal_2003c,
Stasinska_etal_2006}, hereafter K01, K03 and S06, respectively).
Galaxies above these dividing lines have their collisionally excited
lines (\oiii, \nii, etc.)  stronger with respect to recombination
lines (\Ha, \Hb) than SF galaxies, signaling photoionization by a
radiation field harder than that produced by massive young stars.

The right wing is populated by galaxies with Seyfert-like or
LINER-like spectra\footnote{Throughout the paper we will use the words
Seyfert and LINER for galaxies with Seyfert-like and LINER-like
spectra, respectively, regardless of whether or not they are dominated
by non-stellar nuclear activity.}.  Recently, taking advantage of the
huge number of galaxy spectra available in the SDSS, \citet[hereafter
K06]{Kewley_etal_2006} identified a split of the right wing into
Seyfert and LINER branches, better seen in the \oiii/\Hb-\sii/\Ha\ and
\oiii/\Hb-\oi/\Ha\ diagrams, but also visible in the BPT. The physics
behind this dichotomy, and indeed of right wing sources as a whole, is
far less understood than that behind the SF wing.  Moreover, as argued
by \citet{Stasinska_etal_2008} observational selection effects play a
key role in shaping the right wing, and both stellar and non-stellar
ionizing sources can be present.

Notwithstanding these interpretational caveats, for both physical and
practical reasons the BPT diagram has been the main workhorse for
emission line classification for nearly three decades.  The basic
requirement for a reliable BPT classification is that \Hb, \oiii, \Ha
and \nii are {\em all} detected above some minimum signal-to-noise
ratio ($SN_\lambda$). SDSS papers usually adopt a uniform $SN_\lambda
\ge 3$ cut (K03; \citealp{Brinchmann_etal_2004, Li_etal_2006}). This
quality control charges a large toll in terms of number of excluded
objects.  The scale of this problem is often overlooked (see however
\citealp{Miller_etal_2003, Best_etal_2005,Hao_etal_2005}). As shown
here, for the SDSS as a whole, about 1 in 3 emission line galaxies
have \Hb\ and/or \oiii below this threshold.  More worryingly, the
overwhelming majority of these excluded galaxies belong to the right
wing, where close to 2 in every 3 sources suffer from line weakness.

Clearly, no quantitative nor qualitative picture of emission line
galaxies in the local Universe can be complete ignoring weak line
galaxies (WLGs). This is the basic motivation behind this work, whose
main goal is to rescue this numerous, yet often forgotten population
of galaxies from the classification limbo. The physical nature of WLGs
will be discussed in a forthcoming communication.

This paper is structured as follows.  After presenting the sample and
basic data-processing steps (Section \ref{sec:data}), we assess the
size of the WLG population in the SDSS and define sub-types of WLGs
based on which lines prevent a reliable spectral classification
(Section \ref{sec:WLG_stats}).  In Section \ref{sec:DDforWLGs} we
propose alternative diagnostics diagrams which help in placing WLGs in
the standard framework of emission line categories.  Section
\ref{sec:practical_classif} presents an objective method to transpose
the most popular SF/AGN and Seyfert/LINER dividing lines to our more
economic diagnostic diagrams.  These new and more inclusive diagrams
allow the classification of WLGs, leading to a revised census of
emission line galaxies in the nearby Universe, presented in Section
\ref{sec:Application}, which also discusses how WLGs affect the
Seyfert/LINER dichotomy. Section \ref{sec:Conclusions} summarizes our
conclusions.  Hurried readers in search of spectral classification
criteria may want to jump straight to Section 5.6, where equations to
separate SF from AGN and Seyferts from LINERs are presented.


\section{Data}
\label{sec:data}

\subsection{Sample selection}
\label{sec:sample}

The data used in this study come from the 7$^{\rm th}$ data release of
the SDSS \citep{Abazajian_etal_2009}. We start from a raw sample of
926246 galaxies analyzed with the {\sc starlight} code
\citep{CidFernandes_etal_2005}, and apply an initial cut to objects
within the SDSS Main Galaxy Sample \citep{Strauss_etal_2002}. This
leaves nearly 700 thousand galaxies, and already excludes objects with
broad emission lines. A final sample of $\sim$ 370 thousand galaxies
is culled from this list by applying the following criteria.

Since we are interested in objects with weak emission lines, we
removed all spectra where artifacts like bad pixels, imperfect
sky-subtraction or lack of data prevent a clean measurement of any of
the following lines: \oii, \Hb, \oiii, \Ha and \nii. In practice, we
require no faulty pixel within $\pm 15$ \AA\ of these lines. This
guarantees that when any one of these lines is not detected it is
because it is really immersed in the noise or altogether absent,
instead of due to some technical problem.  This conservative cut alone
imposes a substantial (42\%) reduction on the sample.  Few galaxies
outside the 0.024--0.17 redshift interval survive this ``clean lines
only'' cut, so, to round up numbers, we further exclude galaxies
outside this range.  We further trimmed the sample by requiring a
lower limit of 10 for the signal-to-noise ratio in the 4730--4780 \AA\
continuum.  This is done to ensure that enough signal is present to
allow a meaningful stellar population fit, necessary for the
measurement of emission lines, particularly weak ones.

These criteria lead to a sample of 371084 galaxies, hereafter our
``main sample''.  The restriction to good fluxes around the
wavelengths of the main emission lines introduces some peculiarities,
like $z$ gaps when \oiii and \Hb move into the region around the 5577
\AA\ sky line.  Since all demographic arguments will be restricted to
within the sample, such peculiarities do not pose a problem.

None of our selection criteria favors the inclusion of emission line
systems, yet {\em the overwhelming majority of galaxies in our sample
do present emission lines}. For instance, 82\% of them have at least
one of \Ha or \nii with $SN_\lambda \ge 3$. This highlights the fact
that emission lines are nearly ubiquitous, and the importance of
understanding their origin.

\subsection{{\sc starlight} processing and emission line measurements}
\label{sec:starlight_and_elines}

Spectra for all galaxies were processed with version 05 of the
spectral synthesis code {\sc starlight}, which performs pixel-by-pixel
fits of the stellar continuum, delivering a long list of physical
properties, as well as pure emission spectra from which emission line
measurements are performed \citep{CidFernandes_etal_2005}.  The code
itself and its results for the entire SDSS-DR5 database are available
in a VO-like environment at www.starlight.ufsc.br, and products for
the whole DR7 will become available shortly.

As in previous papers in this series, the spectral fits are based on
the \citet{Bruzual_Charlot_2003} models with the STELIB library
\citep{LeBorgne_etal_2003}, ``Padova 1994'' tracks
\citep{Bertelli_etal_1994} and \citet{Chabrier_2003} initial mass
function. These models are being superseded by a new vintage of
evolutionary synthesis calculations incorporating improvements in the
stellar tracks and spectral libraries, which should affect galaxian
properties derived from spectral synthesis codes like {\sc starlight}.
These new models also provide measurably better spectral fits, leading
to differences in emission line measurements.  The differences are not
large, but may be significant for intrinsically weak lines. A full
assessment of these effects will have to await the release of these
new models, but experiments based on preliminary versions indicates
that the main results of this study remain valid \citep{Gomes_2009}.

A difference with respect to the latest papers of the SEAGal
collaboration \citep{CidFernandes_etal_2007, Asari_etal_2007,
Stasinska_etal_2008, ValeAsari_etal_2009}, which used DR5 data, is
that we are now working with DR7.  In addition to the increase by
$\sim 60\%$ in the number of galaxies with respect to DR5, there have
been changes in the reduction pipeline which propagate to differences
in the amplitude and shape of the spectra
\citep{Adelman_McCarthy_et_al_2008,Abazajian_etal_2009}.

Emission lines are measured fitting gaussians to the residual spectrum
obtained after subtraction of the {\sc starlight} fit. The gaussians can
have different widths and offsets, with constraints imposed on lines
from similar ionization levels. Details on this procedure are given
in \citet{Stasinska_etal_2006}.

As noted by \citet{Asari_etal_2007}, the \citet{Bruzual_Charlot_2003}
models have a low amplitude hump in the $\sim 100$ \AA\ interval
around \Hb, such that \Hb emission often sits in a slightly negative
valley in the observed minus model spectrum (an artifact of the STELIB
library, which disappears when using models based on the MILES-library
of \citealp{Sanchez-Blazquez_etal_2006}).  \citet{Asari_etal_2007}
found this effect to be unimportant for their \Hb measurements, but
their analysis focused on SF galaxies, which tend to have strong \Hb.
Here it has a larger impact, since we are specifically interested in
galaxies with intrinsically weak \Hb. To alleviate this problem, the
\Hb flux measurements were performed with respect to a local
``continuum'' in the residual spectrum. No such systematic residual is
found for other emission lines, whose measurement does not require the
extra care devoted to \Hb.


\section{The population of galaxies eliminated by line quality cuts}
\label{sec:WLG_stats}

Throughout this paper a galaxy is said to be an emission line galaxy
(ELG) if both \Ha and \nii are detected with a signal to noise of 3 or
better. Out of the 371084 galaxies in our main sample, 280495 (76\%)
match this definition. Many of these, however, have \Hb and \oiii data
which would be considered unusable by most standards. This section
quantifies this population and introduces the WLG notation used in
later sections.

\subsection{The dramatic effect of $SN_\lambda$ selection}
\label{sec:dramacuts}

\begin{figure*}
\includegraphics[bb= 40 440 570 705,width=0.95\textwidth]{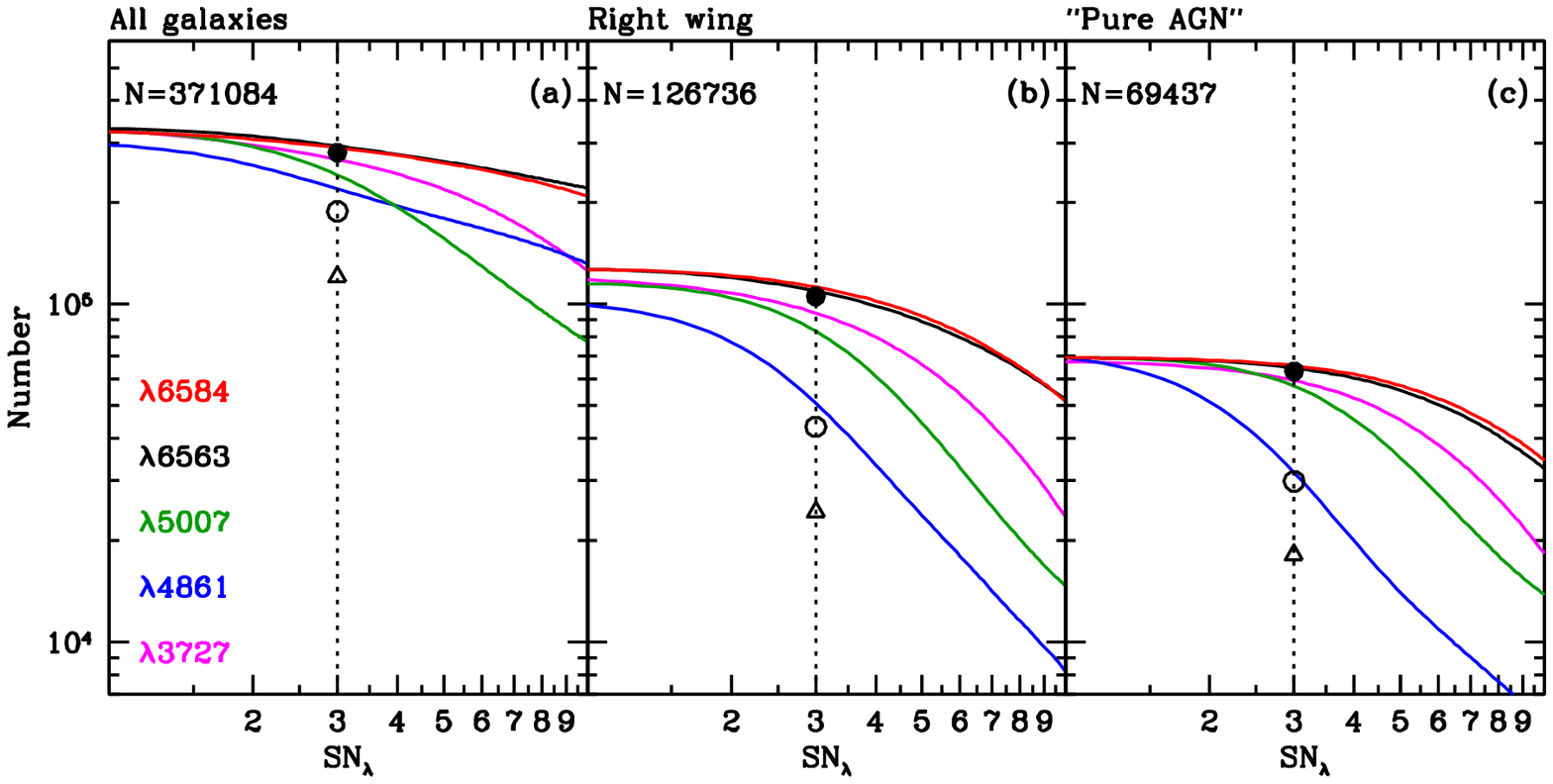}
\caption{Cumulative distributions of signal-to-noise ratio for \oii,
\Hb, \oiii, \Ha and \nii for our main sample (see Section
\ref{sec:sample}). {\em (a)} All galaxies. {\em (b)} Only those with
$\log \nii/\Ha > -0.2$ (right wing sources in the BPT diagram).  {\em
(c)} Only galaxies above the K01 ``extreme starburst'' line in the BPT
plane, often dubbed ``pure AGN'' (but see Section
\ref{sec:MeaningOfDivLines}). In all panels a filled circle marks the
number of galaxies with $SN_{\Ha}$ and $SN_{\nii} \ge 3$ (our
definition of Emission Line Galaxy), while the empty circle
corresponds to $SN_\lambda \ge 3$ in all four BPT lines (called
``Strong Line Galaxies'' in this paper), and a triangle marks the
number of galaxies with \Hb, \oiii, \Ha, \nii plus \oi and the \sii
lines detected at $SN_\lambda \ge 3$. Notice the huge effect of
imposing a $SN_\lambda \ge 3$ cut on lines other than \Ha and \nii,
particularly for non SF galaxies (panels b and c).}
\label{fig:S2N_distribs}
\end{figure*}

A first assessment of the dimension of the WLG population can be
obtained from the distributions of $SN_\lambda$ for the main optical
emission lines.  \citet{Brinchmann_etal_2004} have carried out such an
analysis for their DR1 data; in particular, their Fig.\ 2 shows
histograms of $SN_\lambda$ for the main lines. Our results are shown
in Fig.~\ref{fig:S2N_distribs}, where we plot the cumulative
$SN_\lambda$ distribution for \oii and the four BPT lines.

The left panel shows that \Hb and \oiii are the weakest of the BPT
lines for the main sample as a whole.  Of the 280495 galaxies which
satisfy our definition of ELG, i.e., $SN_{{\rm H}\alpha}$ and
$SN_{[\rm{N}\,\textsc{ii}]} \ge 3$ (marked with a filled circle in the
plot), only 188052 also have $SN_{{\rm H}\beta}$ and
$SN_{[\rm{O}\,\textsc{iii}]} \ge 3$, as indicated by the open
circle. The responsibility for this 33\% global reduction is
approximately equally shared among \Hb and \oiii.

It is crucial to realize that galaxies excluded by a uniform
$SN_\lambda$ cut are {\em not} just a random population which would
spread evenly among strong line sources in the BPT diagram had their
spectra been collected with better signal. To illustrate this,
Fig.~\ref{fig:S2N_distribs}b shows the $SN_\lambda$ cumulative
distributions for the subset of galaxies with $\log \nii/\Ha > -0.2$,
a criterion which completely excludes left wing sources (S06). The \Hb
curve is now well below the one for \oiii for any value of
$SN_\lambda$.  Of the 105414 ELGs in this plot, 74\% also have
$SN_{[\rm{O}\,\textsc{iii}]} \ge 3$, but only 47\% have $SN_{{\rm
H}\beta} \ge 3$. Requiring {\em both} \Hb\ and \oiii\ to have
$SN_\lambda \ge 3$ reduces the sample to 41\% (open circle). Clearly,
weak or undetected \Hb\ emission is the main culprit for this dramatic
(factor of 2.4) drop in numbers of galaxies in the right wing.  A
caveat with this plot is that the $\nii/\Ha$ ratio, used to select
right wing sources, is computed for all galaxies where both lines are
``detected'', ie., $SN_\lambda \ge 1$. For the lowest $SN_\lambda$
values, this ratio is highly uncertain, but since over 83\% of the
sources in Fig.~\ref{fig:S2N_distribs}b have $SN_{\nii}$ and $SN_{\Ha}
\ge 3$, contamination by non-right-wing sources is small.

Fig.~\ref{fig:S2N_distribs}c repeats this numerology including only
objects above the K01 ``extreme starburst'' line in the BPT plane,
leaving only sources normally interpreted as ``pure AGN'', i.e, AGN
whose lines are little or not contaminated by SF (but see section
\ref{sec:MeaningOfDivLines} for a criticism of this reading).  In this
case, just 43\% of the objects have $SN_\lambda \ge 3$ in all 4 lines
involved in the selection process, and thus the results should be
taken as indicative only. Caveats aside, the general message conveyed
by this plot is the same as that spelt by
Fig.~\ref{fig:S2N_distribs}b, namely that low $SN_\lambda$ emission
lines occur more frequently among AGN-like galaxies.

Finally, in all panels of Fig.~\ref{fig:S2N_distribs} a triangle marks
the result of requiring 3 sigma or better detections of all BPT line
plus the \oi and \sii lines, all of which are all explicitly used in
the K06 Seyfert/LINER classification scheme. Clearly, the implied
exclusion factors are prohibitively large.

To summarize:

\begin{enumerate}
\item WLGs comprise a large ($\sim 1$ in 3) fraction of ELGs in the
SDSS.

\item Line weakness is {\em much more severe in the right wing} of the
BPT diagram, where nearly half of the sources fail a $SN_\lambda \ge
3$ cut for all the BPT lines.

\item Further requiring good \oi and \sii measurements imply $\sim 2$
times larger exclusion factors.
\end{enumerate}

\subsection{Weak line galaxies of different kinds: Definitions}

We define a WLG as a galaxy whose \Ha and \nii lines are both detected
with $SN_\lambda \ge 3$, but either or both of \Hb and \oiii have
lower $SN_\lambda$.  In other words, a WLG is an ELG with weak \Hb
and/or weak \oiii.  Conversely, galaxies where all 4 BPT lines have
$SN_\lambda \ge 3$ will be called ``strong line galaxies'' (SLG).
WLGs and SLGs comprise 33\% and 67\% of our ELG sample, respectively
(but recall from Fig.~\ref{fig:S2N_distribs} that these proportions
are $\sim$ inverted in the right wing).

With this definition, WLGs fall into one of 3 possible kinds:

\begin{itemize}
\item WL-H: Weak \Hb\ ($SN_{{\rm H}\beta} < 3$) but strong \oiii\
  ($SN_{[\rm{O}\,\textsc{iii}]} \ge 3$);
\item WL-O: Weak \oiii\ ($SN_{[\rm{O}\,\textsc{iii}]} < 3$) but strong
 \Hb\ ($SN_{{\rm H}\beta} \ge 3$);
\item WL-HO: Weak \Hb\ ($SN_{{\rm H}\beta} < 3$) and \oiii\
($SN_{[\rm{O}\,\textsc{iii}]} < 3$).
\end{itemize}

The WL-H, WL-O and WL-HO denomination is introduced merely to identify
which emission lines are too weak (or missing) to allow a solid
spectral classification. These are {\em not} meant to be interpreted
as new spectral classes. On the contrary, our goal here is precisely
to find out where these objects fit into the current emission line
taxonomical paradigm.

The populations of WLGs of types WL-H, WL-HO and WL-O are comparable
in size: $N_{\rm WL-H} = 38631$, $N_{\rm WL-HO} = 25805$ and $N_{\rm
WL-O} = 28007$.  Note that WLGs include objects where \Hb and/or \oiii
are not detected at all ($SN_\lambda < 1$).  Such non-detections are
genuinely due to line weakness, since we have excluded sources with
faulty pixels around emission lines (see Section \ref{sec:sample}).
\oiii is undetected in 15\% (4288) of WL-Os, while \Hb is undetected
in 25\% (9531) of WL-Hs. Among WL-HOs, 8059 (31\%) have no \Hb, 5773
(22\%) have no \oiii and 1965 (8\%) have neither.  Excluding
non-detections, WL-Hs have median $SN_\lambda$ values of 1.9 in \Hb,
4.5 in \oiii, 7.4 in \Ha, and 7.6 in \nii.  For WL-HOs, $SN_{\Hb} =
1.7$, $SN_{\oiii} = 2.1$, $SN_{\Ha} = 6.0$, and $SN_{\nii} = 5.7$,
while for WL-Os, $SN_{\Hb} = 6.0$, $SN_{\oiii} = 2.3$, $SN_{\Ha} =
24.8$, and $SN_{\nii} = 13.4$.

We point out that our definition of ELG leaves out 22211 sources where
only one of \Ha and \nii satisfies the $SN_\lambda \ge 3$ limit.
There is little one can do in such cases, even if this population
contains some true emission line objects.  Not surprisingly, however,
nearly all (96\%) of these galaxies are also weak in either or both of
\Hb and \oiii: 62\% of them would match our definition of WL-HOs, 30\%
would fall in the WL-H class, but only 4\% would be WL-Os. These
objects will remain excluded from the analysis that follows, but the
numbers above suggest that they can be grouped with WLGs of types H
and HO.

\subsection{Weak line galaxies: Equivalent widths}

\begin{figure}
\includegraphics[bb= 60 170 330 600,width=0.45\textwidth]{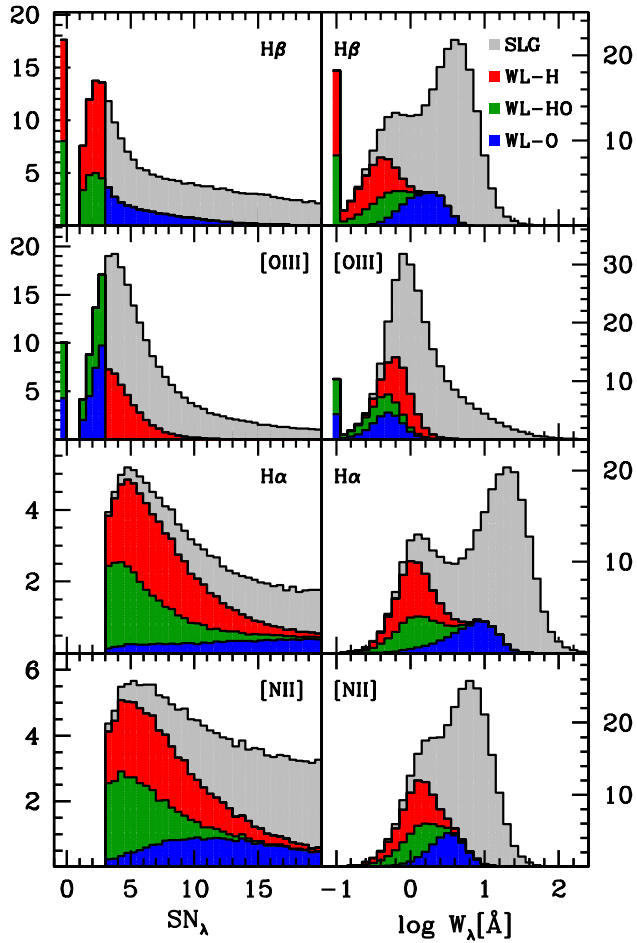}
\caption{Histograms (in thousands of galaxies per bin) of emission
line signal-to-noise ratios (left) and equivalent widths (right) in
the ELG sample.  Red, green and blue regions correspond to WLGs of
types H, HO and O, respectively, while SLGs are painted in grey.
Non-detections ($SN_\lambda < 1$) are grouped in the first bin.}
\label{fig:RefFigSNandEWs}
\end{figure}

We use the term ``weak'' for lines with $SN_\lambda < 3$, a
terminology which would be innapropriate in cases where a line with a
large equivalent width ($W_\lambda$) is immersed in a noisy continuum
(resulting in a low $SN_\lambda$).  For the sample considered here,
however, low $SN_\lambda$ does correspond to low $W_\lambda$. This is
illustrated in Fig.\ \ref{fig:RefFigSNandEWs}, where we show the
distributions of $SN_\lambda$ and $W_\lambda$ for the four lines
involved in our definitions of WLGs.  The top panels show that WL-Hs
are the sources with the smallest \Hb equivalent widths, with a median
$W_{\Hb}$ value of just 0.3 \AA.  Similarly, WL-Os (blue areas in
Fig.\ \ref{fig:RefFigSNandEWs}).  have low $W_{\oiii}$, while
$W_{\Hb}$ and $W_{\oiii}$ are both low in WL-HOs (green areas).

Importantly, as seen in the bottom panels of Fig.\
\ref{fig:RefFigSNandEWs}, WL-Hs are also among the sources in the low
end of the \oiii, \Ha and \nii equivalent width distributions, even
though, by definition, $SN_\lambda \ge 3$ in all these lines. In the
median, $W_{\oiii} = 0.7$, $W_{\Ha} = 1.1$, and $W_{\nii} = 1.2$ \AA\
for WL-Hs.  These galaxies thus have low $W_\lambda$ in {\em all}
major optical emission lines. Similar comments apply to WL-HOs. The
requirement of $SN_{\Hb} \ge 3$ therefore biases SLG samples {\em
against} objects whose emission lines in general have low
$W_\lambda$. Equivalent widths are not considered in current emission
line classification schemes, but it is well known that low $W_\lambda$
systems tend to avoid the tips of the AGN and SF wings in the BPT
diagram. One thus expects to find few Seyferts and low metallicity SF
galaxies among WL-Hs and WL-HOs.

The situation for WL-Os is somewhat different, in the sense that, as
expected from the $SN_{\oiii} < 3$ condition, they have low
$W_{\oiii}$, but the equivalent widths of other emission lines are not
as small as those in WL-Hs and WL-HOs. For instance, the median
$W_{\Ha}$ is about six times larger in WL-Os than in other WLGs,
despite some overlap in the distributions (see bottom-right panel in
Fig.\ \ref{fig:RefFigSNandEWs}).

Given that $SN_\lambda$ and $W_\lambda$ are strongly coupled, one may
wonder how important aperture effects are in shaping the WLG
population. The fraction of WLGs over ELGs increases from $\sim 15\%$
at $z = 0.024$ to a little short of 60\% at $z = 0.17$. A naive
interpretation of this trend is that as $z$ increases more galaxy
light enters the SDSS fiber, so nuclear emission lines become
increasingly diluted and harder to detect. This suggests that, besides
intrinsic line weakness, aperture effects play an important role in
defininig which galaxies have weak lines and which do not. While this
effect is certainly present, it is heavily convolved with other
potentially more important distance dependencies. Stellar masses, for
instance, also grow strongly with $z$. We shall defer an analysis of
these issues to a forthcoming paper.


\section{Alternative Diagnostic diagrams}
\label{sec:DDforWLGs}

The previous section has shown that WLGs comprise a large fraction of
ELGs in the SDSS. Clearly, this numerous population deserves
attention.  We now look for ways to classify ELGs which include this
usually forgotten population, trying to place WLGs in the framework of
standard emission line categories.

\subsection{The BPT diagram}
\label{sec:BPT}

\begin{figure}{}
\includegraphics[bb= 30 170 340 690,width=0.494\textwidth]{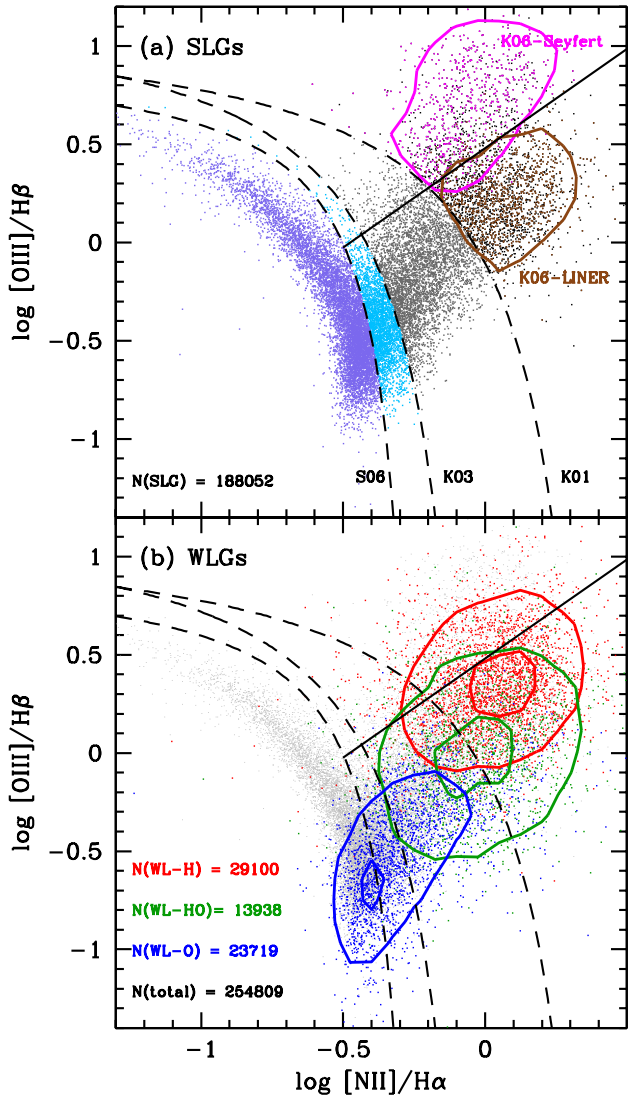}
\caption{{\em (a)} The BPT diagram for Strong Line Galaxies. The
dashed curves represent the SF/AGN division lines from S06, K03 and
K01. Colors are used to indicate spectral classifications according to
these lines. Magenta and brown points are Seyferts and LINERs
according to the K06 criteria, and contours with these same colours
correspond to a number density of 20\% of the peak value.  The
straight line is the Seyfert/LINER division line proposed in this
paper (equation \ref{eq:TranspDivLine_K06_BPT}). {\em (b)} BPT diagram
for Weak Line Galaxies, ie., those where either or both of \Hb and
\oiii fail a $SN_\lambda \ge 3$ cut.  Red, green and blue points and
contours mark the location of WLGs of types WL-H (ie., $SN_{\Hb} < 3$
and $SN_{\oiii} \ge 3$), WL-HO ($SN_{\Hb}$ and $SN_{\oiii} < 3$), and
WL-O ($SN_{\Hb} \ge 3$ and $SN_{\oiii} < 3$), respectively. For
reference, SLGs (the same as in panel a) are shown as the light grey
points in the background.  Contours correspond to number densities of
20 and 80 \% of the peak density. Note: For clarity, in these and all
other diagrams in this paper {\em only one in every ten galaxies is
plotted.}  {\ATT
 The actual numbers of galaxies in each category are
listed in the lower left corner of the panels.  N(total) is the number
of SLG + WLG in the corresponding diagram.
 ~~~~ ~~~~ ~~~~ ~~~~ ~~~~ ~~~~ ~~~~ ~~~~ ~~~~ ~~~~ ~~~~
 ~~~~ ~~~~ ~~~~ ~~~~ ~~~~ ~~~~ ~~~~ ~~~~ ~~~~ ~~~~ ~~~~ .
}
}
\label{fig:WLGsOnBPT}
\end{figure}

Fig.\ \ref{fig:WLGsOnBPT}a shows the BPT diagram for the entire sample
of SLGs defined in Section \ref{sec:data}. As in all other diagnostic
diagrams in this paper, for clarity purposes only one in every ten
galaxies is plotted, while the actual number of galaxies in each
category which could have been plotted is given in the legend. None of
the line ratios have been corrected for reddening, but their
reddening-dependence is negligible in the BPT. The curves show the
SF/AGN border-lines defined by S06, K03 and K01, and points are
color-coded accordingly. Points above the K01 line, often interpreted
as ``pure AGN'', are plotted in magenta or brown if they match the K06
criteria for Seyferts and LINERs, respectively. Points in black in
this same region cannot be classified with their criteria due to the
weakness of one or both of \oi and \sii, or due to conflict between
the classification obtained in different diagnostic diagrams (the
``ambiguous'' category of K06). The straight line marks the
Seyfert/LINER division line derived in Section
\ref{sec:Sey_LINER_div_lines}, proposed to fix this ambiguity.

Even though WLGs have (by definition) unreliable \oiii and/or \Hb line
intensities, it is instructive to estimate their location in the BPT
diagram.  This is done in Fig.\ \ref{fig:WLGsOnBPT}b, where we
transgress the common sense rule of using only good line measurements
by plotting WL-H (in red), WL-HO (green), and WL-O (blue) sources.
Contours with these same colors help tracing the location of these
different types of WLGs in the BPT diagram.  For reference, the
background points represent the same SLGs as panel a. Counting both
SLGs and WLGs, panel b includes N(total) $= 254809$ sources.

Given the small difference in wavelength between \oiii and \Hb, the
relation between $SN_{\oiii}$ and $SN_{\Hb}$ should be similar to that
between the \oiii and \Hb fluxes, and thus one would expect WLGs of
types H, HO and O to have $\oiii/\Hb > 1$, $\sim 1$ and $< 1$,
respectively.  These expectations are fully confirmed by Fig
\ref{fig:WLGsOnBPT}b. WL-Hs have $\log \oiii/\Hb = +0.34 \pm 0.16$ and
$\log \nii/\Ha = +0.02 \pm 0.12$ (median $\pm$ semi interquartile
range).  They therefore lie within the region usually associated with
``pure AGN''. Most of them sit on the lower side of the right wing,
the realm of LINERs.  WL-HOs have $\log \oiii/\Hb = 0.00 \pm 0.19$ and
$\log \nii/\Ha = -0.04 \pm 0.13$, which place them well within the
right wing, heavily overlapping with WL-H galaxies.

One should not be surprised to find that WL-Hs and WL-HOs are
mostly AGN-like. Our definition of ELGs requires $SN_\lambda \ge 3$ in
both \Ha and \nii, such that as one moves from the left to the right
wing $SN_{\Ha}$ becomes the limiting factor. The median value of
$SN_{\Ha}$ drops from 9 at $\log \nii/\Ha = 0$ to 4 for $\log \nii/\Ha
> 0.4$.  With \Ha so close to our $SN_\lambda$ cut, and since $F(\Hb)
< F(\Ha) / 3$, weak \Hb sources are bound to be found in the right
wing.

While essentially no WL-H nor WL-HO galaxy falls in the SF region as
defined by either K03 or S06, most WL-Os are consistent with a SF
classification. They populate the bottom of the left wing, where the
most massive and metal rich SF galaxies are found
\citep{Asari_etal_2007,Stasinska_etal_2008}. Although there is some
overlap with WL-HOs, most WL-Os belong to the left wing, with 76\% of
them having $\log \nii/\Ha < -0.2$.  This sets the bulk of these WLGs
apart from type H and HOs, which are intrinsically right wing sources.

While instructive, these results should be read with care, since
dealing with low $SN_\lambda$ lines involves important biases (e.g.,
\citealp{Rola_Pelat_1994}).  Furthermore, as many as $28 \%$ of our
WLGs are not even not plotted in Fig.~\ref{fig:WLGsOnBPT} because of a
missing \Hb or \oiii measurement. One would therefore like to confirm
these results with more robust alternative diagrams.

\subsection{The BPT$\alpha$ diagram}
\label{sec:BPTHa}

\begin{figure}
\includegraphics[bb= 30 170 340 690,width=0.5\textwidth]{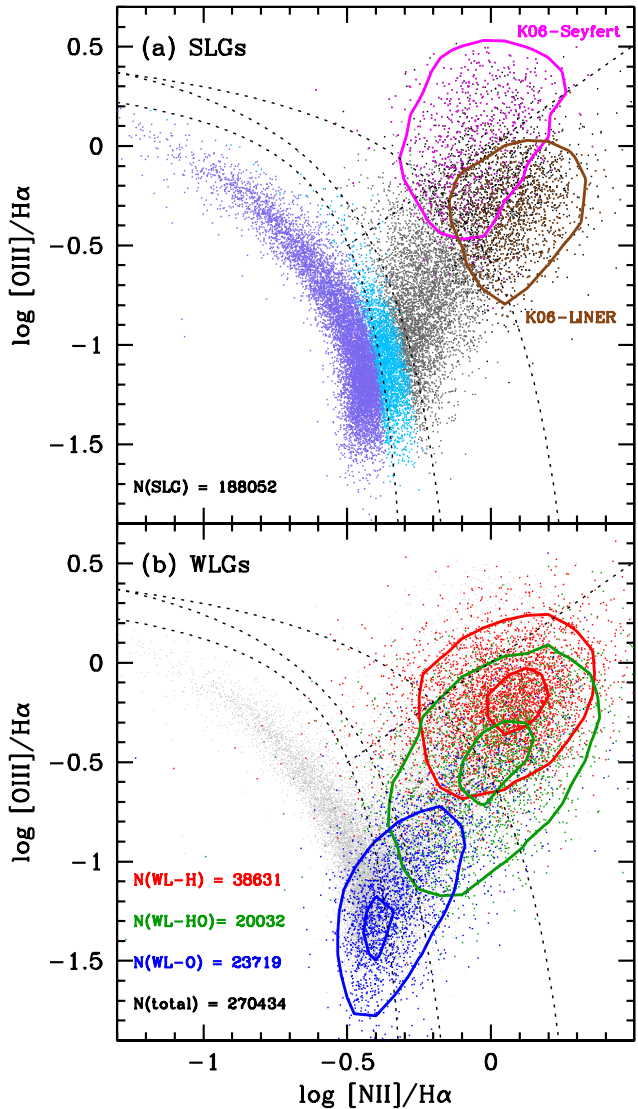}
\caption{The BPT$\alpha$ diagram: As the BPT, but replacing \Hb by
\Ha. Panels a and b are analogous to those in Fig.\
\ref{fig:WLGsOnBPT}, except that in b the number of galaxies is larger
because \Hb is not used.  Note the downwards shift of the data points
with respect to the dividing lines (the same as in Fig.\
\ref{fig:WLGsOnBPT} but shifted by $- \log 3$) because of
extinction. }
\label{fig:WLGsOnBPTa}
\end{figure}

One way to deal with a weak \Hb line is to replace it by a stronger
hydrogen line, \Ha. Fig.~\ref{fig:WLGsOnBPTa} shows the
``BPT$\alpha$'' diagram, whose horizontal axis is the same as in the
BPT diagram but the vertical axis is $\oiii/\Ha$. Panel a shows SLGs
(exactly the same ones appearing in Fig.~\ref{fig:WLGsOnBPT}a). WLGs
are shown in panel b, following the same color coding as in
Fig.~\ref{fig:WLGsOnBPT}b. The dotted lines are the same as in
Fig.~\ref{fig:WLGsOnBPT}a, except for a 0.48 dex downwards shift
corresponding to a dust-free case B \Ha/\Hb ratio of 3.

As expected, the seagull shape is preserved in the BPT$\alpha$
diagram. The relative locations of WL-H, WL-HO and WL-O systems is
also the same as in the original BPT plane, confirming the overall
picture sketched above, namely, that WL-H and WL-HO sources are mostly
LINER-like right wing sources, while WL-O systems behave like
metal-rich SF galaxies.  The advantage of this plot is two-fold:
First, WL-H systems have $SN_\lambda \ge 3$ in all 3 emission lines
involved. Second, it allows the inclusion of 17590 WLGs (19\% of the
whole WLG population) absent from Fig.~\ref{fig:WLGsOnBPT}b because of
lacking \Hb fluxes (non-detections, i.e., $SN_{\Hb} < 1$).  However,
even though in the case of WL-HOs the y-axis is more robust than in
the BPT, this diagram does not solve the problem of low quality \oiii
fluxes, and thus the BPT$\alpha$ is not suitable to classify WL-Os and
WL-HOs except in a statistical sense.

Unlike \oiii/\Hb, the \oiii/\Ha\ ratio is sensitive to reddening.  For
a $R_V = A_V / E(B-V) = 3.1$ \citet*{Cardelli_Clayton_Mathis_1989}
law, 1 magnitude extinction in the V band increases \oiii/\Hb by an
insignificant 0.02 dex, but decreases \oiii/\Ha, by 0.12 dex, causing
detectable downwards shifts in the BPT$\alpha$ plane.  Such shifts can
indeed be seen comparing Figs.~\ref{fig:WLGsOnBPT} and
\ref{fig:WLGsOnBPTa} and using the dividing lines as a reference.
However, given that reddening correlates with other galaxy properties,
both for SF and AGN galaxies (\citealp{Stasinska_etal_2004}; K06), the
BPT$\alpha$ diagram should still provide a meaningful diagnostic to
separate ELGs into classes.  This is confirmed by the clear separation
of points of different colors in Fig.\ \ref{fig:WLGsOnBPTa}a.  Insofar
as classification is concerned, the most worrying class confusion
caused by reddening occurs in the upper half of the right wing, where,
because of their higher reddening (e.g., K06), Seyferts intrude into
the zone occupied by LINERs, resulting in a substantial overlap (see
the contours in Fig.\ \ref{fig:WLGsOnBPTa}a).  Some confusion also
occurs between SF and AGN when using the K01 line, but reddening
causes little confusion of SF and AGN classes as defined by S06 and
K03, given their nearly vertical dividing lines in the region
corresponding to the body of the seagull.

\subsection{The BPTo2 diagram}

\label{sec:o3o2}

\begin{figure}
\includegraphics[bb= 30 170 340 690,width=0.5\textwidth]{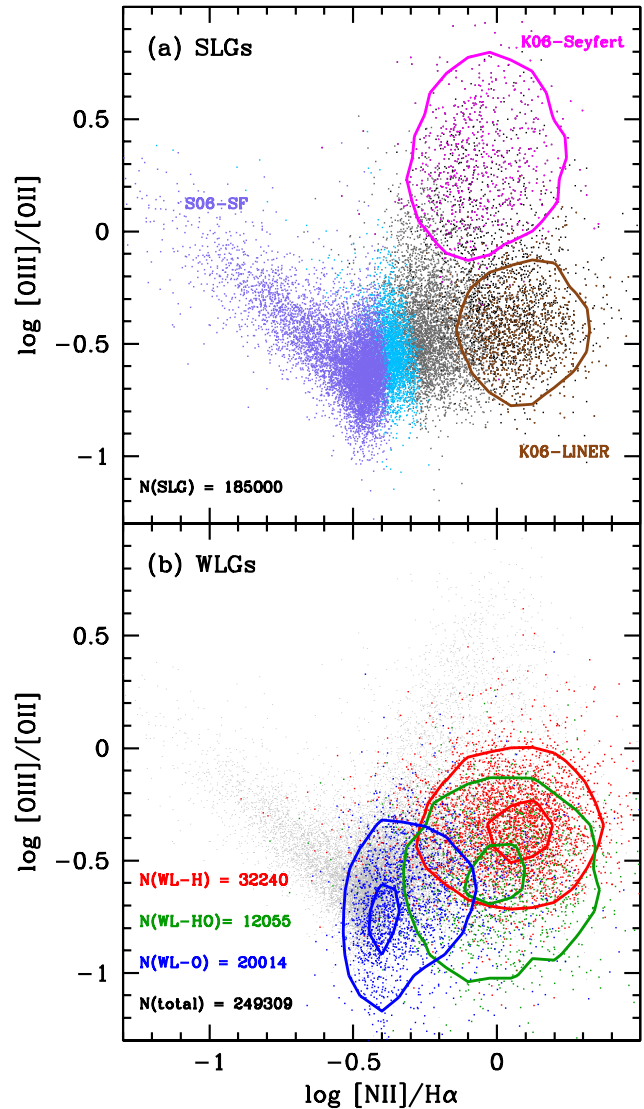}
\caption{The BPTo2 diagram: \oiii/\oii\ vs \nii/\Ha.  Analogous to
Figs.\ \ref{fig:WLGsOnBPT} and \ref{fig:WLGsOnBPTa}, but adding the
requirement of $SN_{\oii} \ge 3$ to both strong and weak line
galaxies.}
\label{fig:WLGsOnBPTo2}
\end{figure}

Another way to display galaxies with weak \Hb is to replace \Hb by
\oii in the BPT diagram. As shown in Fig.~\ref{fig:S2N_distribs}, \oii
is much less affected by low $SN_\lambda$ than \Hb. To include \oii in
the analysis we momentarily modify our definition of ELGs by adding
the requirement that $SN_{\oii} \ge 3$. This implies a 9\% reduction
in the ELG sample as a whole, 2\% in the SLG sample and 24\% of WLGs
as a whole, a modest price to pay in exchange for the replacement of a
bad (or non-existent) datum by a convincing detection.

Fig.~\ref{fig:WLGsOnBPTo2} shows the \oiii/\oii versus \nii/\Ha
diagram (the same as in Fig.~3 of BPT's paper, except for the order of
the axes). Panel a shows galaxies with $SN_\lambda \ge 3$ in all four
BPT lines {\em and} \oii, while WLGs with $SN_{\oii} \ge 3$ are
plotted in panel b.

Like the BPT diagram, this ``BPTo2'' diagram also opens up into SF and
AGN wings. The main difference is that the right wing clearly splits
into Seyfert and LINER branches, an effect which, although present, is
much less pronounced in the BPT and BPT$\alpha$ planes. This split is
evident in the contours corresponding to K06 Seyferts and LINERs
(Fig.~\ref{fig:WLGsOnBPTo2}a), which, unlike in previous diagrams, do
not overlap.  The similarity of WL-H and W-HO sources is even more
evident in terms of BPTo2 coordinates. Both types of WLGs populate the
lower $\oiii/\oii$ branch of the right wing, where LINERs are
located. A negligible number of WLGs intrude into the Seyfert branch,
which is almost exclusively populated by SLGs.

Why does this diagram work so well? In this case, $A_V = 1$ mag
implies an increase of \oiii/\oii of 0.17 dex with respect to its
intrinsic value. The enhanced distinction between Seyferts and LINERs
seen in Fig.\ \ref{fig:WLGsOnBPTo2} is partly due the fact that
ionization state and reddening are positively correlated, i.e, that
Seyferts are more heavily reddened than LINERs
(\citealp*{Ho_Filippenko_Sargent_2003}; K06). Whereas in the
BPT$\alpha$ plane this effect causes a certain degree of confusion of
Seyferts and LINERs, in the BPTo2 it enhances the distinction between
these two classes.  Besides this extrinsic effect, \oiii/\oii\ is more
sensitive to the ionization state than \oiii/\Hb, which also enhances
the separation between Seyferts and LINERs. Moreover, \oii starts
being collisionally de-excited at densities well below the critical
density of \oiii, so a correlation between density and ionization
state would cause a split in the \oiii/\oii ratio. Since the narrow
line region of Seyferts is generally denser than that of LINERs
\citep{Ho_Filippenko_Sargent_2003}, this could be yet another cause
for the marked split seen in Fig.\ \ref{fig:WLGsOnBPTo2}, although we
note that this is a minor effect in our SDSS data, where Seyferts and
LINERs span similar values of the density sensitive ratio between the
6731 and 6716 \sii lines. All these effects act in the same sense,
explaining why the BPTo2 diagram is so effective in separating
Seyferts from LINERs.

The gain in statistics with the BPTo2 diagram is significant,
especially for WL-Hs, 83\% of which are in Fig.\
\ref{fig:WLGsOnBPTo2}.  {\em All} the data for this subset of objects
now rely on convincing ($SN_\lambda \ge 3$) detections, greatly
alleviating the problem of classification in the right wing.  WL-Os
are also well represented (83\% as well), while WL-HOs are present at
a 56\% level, but in both cases the y-axis still contains uncertain
\oiii fluxes. Inevitably, the requirement of $SN_{\oii} \ge 3$ data
makes the BPTo2 somewhat less inclusive than the BPT$\alpha$ diagram,
but the difference is relatively small, and largely compensated by its
much better Seyfert/LINER diagnostic power.

Neither the BPT$\alpha$ nor the BPTo2 allow a robust classification of
WL-HOs and WL-Os. Due to the statistical power of the SDSS, diagrams
utilizing $SN_{\oiii} < 3$ data show an expected pattern, but
individual objects cannot be reliably classified using such uncertain
data. The next section presents an alternative which circumvents this
limitation.

\subsection{The EW$\alpha$n2 diagram}
\label{sec:ewan2}

\begin{figure}
\includegraphics[bb= 45 170 340 690,width=0.5\textwidth]{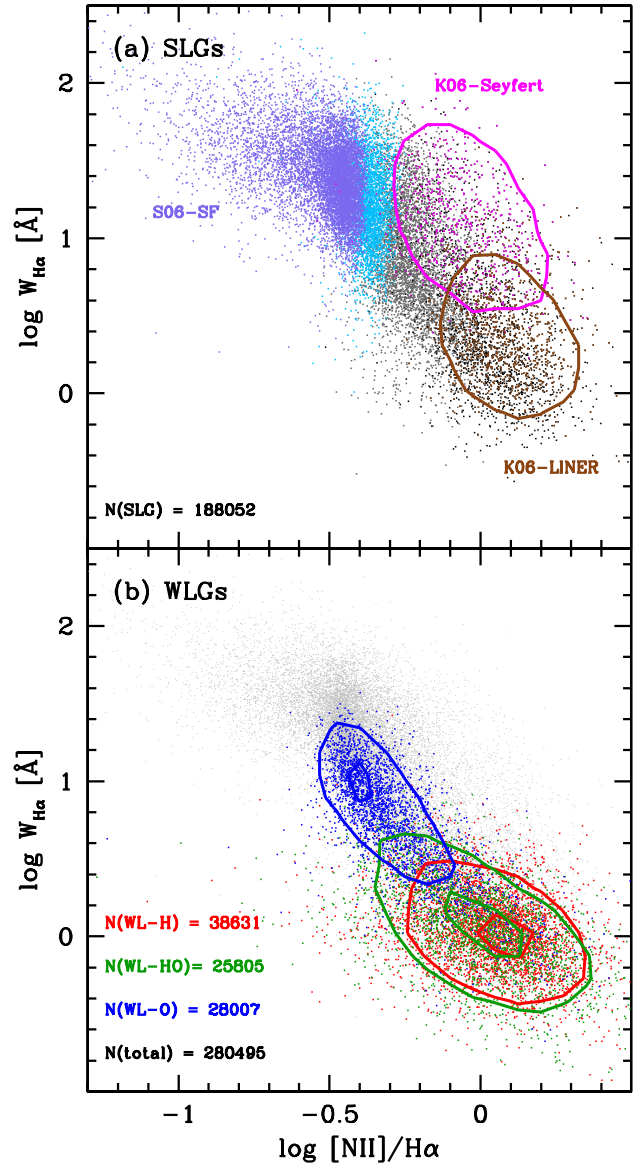}
\caption{The EW$\alpha$n2 diagram: $W_{\Ha}$ vs \nii/\Ha. Colours and
contours as in previous diagnostic diagrams. In panel a only galaxies
with $S/N \ge 3$ in all BPT lines are plotted (SLGs), whereas in panel
b the only requirement is that $S/N \ge 3$ in \Ha and \nii.}
\label{fig:WLGsOnEWhan2}
\end{figure}

The most economic way to classify galaxies is using just two lines.
\Ha\ and \nii\ are the best for this, both from the point of view of
the number of galaxies that can be treated (Fig.\
\ref{fig:S2N_distribs}) and from the point of view of the physical
relevance of the line ratio. \citet{Miller_etal_2003}, 
\citet{Brinchmann_etal_2004} and S06 have already argued for a SF/AGN
classification using the \nii/\Ha\ ratio only. However, such a
classification does not allow one to distinguish Seyferts from LINERs.

We propose to use the equivalent width of \Ha to break this
degeneracy. This proposition entails a radical change in emission line
classification paradigm, in the sense that line ratios and equivalent
widths measure different things.  Emission line ratios trace physical
conditions in the ionized gas, while (neglecting escape of ionizing
photons) $W_{\Ha}$ measures the power of the ionizing agent with
respect to the optical output of the host's stellar population. One
can justify this option on purely heuristic grounds: Seyfert galaxies
are known to have higher values of $W_{\Ha}$ than LINERs, so why not
use this to classify galaxies, especially when no other option is
available?

Fig.~\ref{fig:WLGsOnEWhan2} plots $W_{\Ha}$ versus \nii/\Ha, the
``EW$\alpha$n2 diagram''. Its layout is like that of previous
diagnostic diagrams, with SLGs on the top and WLGs on the bottom.
This is the only diagram that allows us to plot {\em all} our 280495
ELGs. Furthermore, by definition, only $SN_\lambda \ge 3$ data is
used.

\begin{figure}
\includegraphics[bb= 30 170 340 690,width=0.5\textwidth]{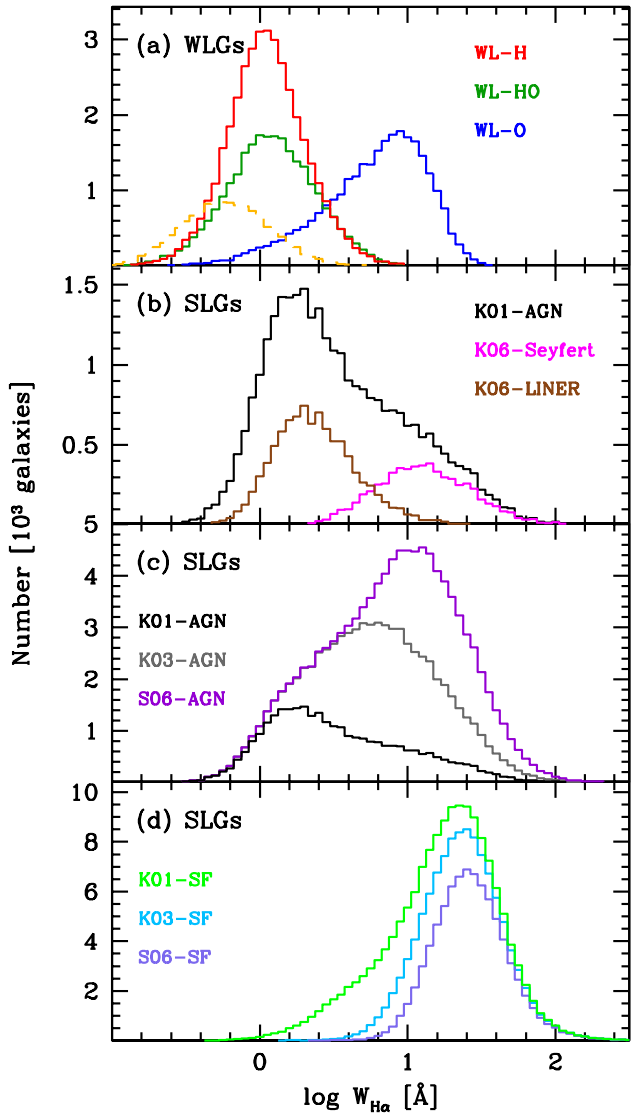}
\caption{Distribution of $W_{\Ha}$ according to K01, K03, K06 and S06
spectral classes (bottom panels) and WLG-type (top).  Panel a shows
exclusively WLGs, whereas panels b, c and d are for SLGs. Panels d and
c show results for the SF and AGN classes in K01, K03 and S06, while
in panel b we show only K01-AGN and the Seyfert/LINER subdivision of
K06. The dashed line in panel a corresponds to galaxies with $SN_{\Ha}
\ge 3$ but which are excluded from out ELG sample because $SN_{\nii} <
3$.}
\label{fig:EWha_distribs}
\end{figure}

The SF/AGN diagnostic power of this diagram resides in the horizontal
axis, while Seyferts and LINERs overlap in \nii/\Ha, but are well
separated in $W_{\Ha}$ (see also Fig.\ \ref{fig:EWha_distribs}).  As
could be anticipated from the morphology of the BPT diagram, SF
galaxies defined by either the S06 or the K03 criteria form nearly
vertical boundaries in the EW$\alpha$n2 diagram, and thus can be well
separated in terms of \nii/\Ha alone. In contrast, SF and AGN systems
defined according to the K01 scheme are hopelessly mixed in this
diagram, with substantial overlaps in both horizontal and vertical
directions.  For the reasons discussed in Section
\ref{sec:MeaningOfDivLines}, this is neither surprising nor a serious
drawback.

Turning to WLGs, Fig.~\ref{fig:WLGsOnEWhan2}b shows that, once again,
WLGs of types H and HO overlap strongly with each other, occupying the
region filled by LINERs in the SLG sample. As in other diagrams, WL-Os
line up with the metal-rich (large \nii/\Ha) SF galaxies, with a tail
of objects stretching towards WL-Hs and WL-HOs.

Fig.\ \ref{fig:EWha_distribs} shows the distribution of $W_{\Ha}$
split by spectral class. Panel d shows the galaxies classified as SF
according to the K01, K03 and S06 dividing lines on the BPT diagram,
whereas panel c shows the corresponding AGN histograms.  Panel b
repeats the K01-AGN histogram, this time also showing the split into
Seyfert and LINER subtypes according to the K06 scheme, which is based
on 7 emission lines and 3 diagnostic diagrams. (As already mentioned
in Section \ref{sec:BPT}, not all objects above the K01 line in the
BPT diagram can be classified in the K06 Seyfert/LINER scheme, which
explains why the magenta and brown lines do not add up to the black
one). All these panels are for SLGs only.

Panel a in Fig.\ \ref{fig:EWha_distribs} shows our WLGs, confirming
that, despite some overlap, WL-Os have typical $W_{\Ha}$ values almost
a full order of magnitude larger than WL-Hs and WL-HOs, and that the
latter two are indistinguishable in terms of $W_{\Ha}$.  The median
$\pm$ semi interquartile ranges of $W_{\Ha}$ are $6.7 \pm 3.5$, $1.1
\pm 0.4$, and $1.2 \pm 0.6$ \AA\ for WL-O, WL-H and WL-HO,
respectively.  K06-LINERs in the SLG sample (Fig.\
\ref{fig:EWha_distribs}b) have a $W_{\Ha}$ distribution which overlaps
with those of WL-Hs and WL-HOs, but is skewed towards somewhat larger
values: $W_{\Ha} = 2.2 \pm 1.0$ \AA. This difference is most likely
due to the more stringent requirements for a full K06 classification,
which, by requiring good detections in many more lines, ends up
skewing $W_{\Ha}$ towards larger values and indirectly excluding
objects which are otherwise similar.


\subsection{Summary}

All the diagrams studied in this section point to the following:

\begin{enumerate}

\item Emission line galaxies with weak \Hb and those with weak \Hb and
\oiii (WL-H and WL-HO) are predominantly LINERs.

\item Part of the WL-HO population straddles the regions between bona
fide LINERs and metal-rich SF galaxies in diagnostic diagrams.
Objects in such intermediate locations are usually
called ``composite'' in current taxonomy.

\item Galaxies with weak \oiii (WL-O) are predominantly metal rich SF
galaxies, though some intrude into the LINER zone of diagnostic
diagrams.

\item Few WLGs are Seyferts.

\end{enumerate}

Regarding the alternative diagnostic diagrams proposed in this section
and their ability to rescue WLGs from the spectral classification ``no
man's land'', we have seen that:

\begin{enumerate}

\item \Hb can be replaced by \Ha or by \oii without significant loss
of classification power. This solves the problem of weak \Hb sources
(about 41\% of all WLGs), which can all be appropriately classified on
the basis of $SN_\lambda \ge 3$ lines exclusively.

\item Seyferts and LINERs are much better distinguished in the BPTo2
diagram than in either the BPT or the BPT$\alpha$.

\item \nii/\Ha does a reasonable job in separating SF from AGN as
defined by S06 and K03, but does not separate K01-SF from K01-AGN.

\item The equivalent width of \Ha is the most economic way of
distinguishing Seyferts from LINERs in the SDSS. 

\end{enumerate}

\section{Transposition of standard  classification schemes
to our alternative diagrams}
\label{sec:practical_classif}

We have shown in the preceding section that a reasonably consistent
emission line classification of most SLGs and WLGs can be achieved in
various diagnostic diagrams. We now convert these results into
equations which allow one to transpose standard SF/AGN and
Seyfert/LINER dividing lines to diagnostic diagrams other than those
they were originally based on.

It is important to emphasize that the new dividing lines presented
below are mere {\em transpositions} of other people's classification
criteria. Specifically, we transpose the widely used K01, K03 and S06
SF/AGN border lines, and the K06 Seyfert/LINER division to the more
economic diagnostic diagrams discussed above. We are {\em not},
therefore, introducing independent classification schemes in a field
were they already abound. This transformation is achieved with an
adaptation of the optimal class separation technique of \citet[see
also \citealp{Mateus_etal_2006}]{Strateva_etal_2001}.  Although our
main motivation is to provide practical criteria to classify WLGs, the
results below are useful to ELGs in general.

Our objective transposition methodology does not overcome the
limitations and ambiguities of spectral classification based on
emission lines. Deficiencies in the reference classification schemes
are propagated to the more widely applicable border lines derived
below. Users of such classification schemes should be fully aware of
such caveats. This is why, before presenting our results, we open a
``parenthesis'' to ponder which of the three classification schemes
studied here best reflects the fundamental distinction between SF and
AGN galaxies in the BPT diagram.

\subsection{Overtones of emission line classification schemes}
\label{sec:MeaningOfDivLines}

There has been some ambiguity, over the years, in the separation
between SF and AGN galaxies in the BPT diagram. A now widely used
scheme is to consider that all galaxies below the K03 line are ``pure
SF'' systems and all galaxies above the K01 line are ``pure AGN'',
while those in between are dubbed ``composite''. This qualitatively
plausible terminology is, however, misleading and inconsistent with
both models and data.

The K01 line was originally designed to select, from a sample of
galaxies, those that certainly harbor an active black hole. Their
``extreme starburst'' line in the BPT plane was obtained by
considering the upper envelope of model-nebulae ionized by massive
stars, considering a wide range of parameters and several stellar
population synthesis models. Hence, according to the K01 models,
sources currently classified as composites (i.e. between the K01 and
K03 lines) do not require the presence of an AGN, and, conversely,
locations above the ``extreme starburst'' line may well be reached by
composite SF+AGN systems.  The K01 line in the BPT plane was never
intended to trace the frontier of ``pure AGN'', as it is used
nowadays.  On the contrary, its goal was to define a {\em lower}
boundary for SF+AGN composites.

It is also fit to recall that photoionization models with either a
pure AGN \citep{Ferland_and_Netzer_1983,
Halpern_and_Steiner_1983,Stasinska_1984} or a purely old stellar
population \citep{Stasinska_etal_2008} are able to cover the region
between the K03 and K01 lines without mixing massive stars and AGN at
all. From a more empirical point of view, stellar population studies
(\citealp{Schawinski_etal_2007}; \citealp{CidFernandes_etal_2008})
show that galaxies with no ongoing star-formation can populate this
zone of the BPT diagram.  Clearly, these systems are not truly SF+AGN
composites.

All in all, the use of the term ``composite'' to denote objects
between the K01 and K03 lines is misleading, even if part of the
galaxies in this zone are indeed mixtures of SF and AGN.  If anything,
such sources should be called ``intermediate'', i.e., in between the
bottom and the tip of the right wing. 

Besides these interpretational issues, a more serious problem with the
K01 line is that, as it became clear with the SDSS, real SF galaxies
fall well below and to the left of it. In fact, the K01 line bears no
resemblance whatsoever to any structure in the observational BPT.
This mismatch prompted K03 to propose a dividing line more connected
to the data. The K03 line was drawn empirically by simply displacing
the K01 line in order to better trace the observed distribution of SF
galaxies in the BPT diagram. This line runs a little higher than the
upper envelope of the bulk of SF galaxies, and, more importantly, it
is somewhat arbitrary at the bottom of the BPT diagram, which hosts a
large proportion of SDSS galaxies. S06 produced a more stringent line
which follows more closely the upper envelope of the bulk of SF
galaxies and was extrapolated to the ``body'' of the seagull, where the
left and right wings meet and no clear frontier is seen. This
extrapolation was achieved by means of photoionization models, and may
obviously be wrong, but it is, for the moment, the best available. 

We thus feel that the S06 divisory line between SF and AGN is better
motivated than the K03 one, although, depending on the problem at
hand, one might prefer using the K03 line, and for sources well within
the right wing it makes no practical difference which of the two lines
is used.  On the other hand, because of its completely artificial
shape in regard with the population of real galaxies in the BPT plane,
the K01 SF/AGN division leads to different results, and to an
ill-defined classification of galaxies into various groups.

Because of it widespread use, we keep the K01 line in the analysis
that follows, but the considerations above show that there are plenty
of reasons to reconsider its role in the spectral classification of
galaxies.

\subsection{SF/AGN border lines in the BPT$\alpha$ and BPTo2 diagrams}
\label{sec:SF_AGN_div_lines}

\begin{table*} 
\caption{Optimal $y = a + b / (x + c)$ SF/AGN dividing lines}
\begin{tabular}{llcclcccccccc}       
\hline\hline 
 diagram  &  line  &  $y$ & $x$ &  $a$ & $b$ & $c$ & ${\cal C}_{\rm SF}$ & ${\cal C}_{\rm AGN}$ &
${\cal R}_{\rm SF}$ & ${\cal R}_{\rm AGN}$ & ${\cal P}$\\
\hline 
 BPT          &  K01 & log \oiii/\Hb  & log \nii/\Ha  &   1.19  & 0.61  & -0.47  & -- & --  & -- &  -- & 1 \\
 BPT          &  K03 & log \oiii/\Hb  & log \nii/\Ha  &   1.30  & 0.61  & -0.05  & -- & --  & -- &  -- & 1 \\
 BPT          &  S06 & log \oiii/\Hb  & log \nii/\Ha  &   0.96  & 0.29  & +0.20  & -- & --  & -- &  -- & 1 \\
\hline 
 BPT$\alpha$  &  K01 & log \oiii/\Ha  & log \nii/\Ha  &  0.69  &  0.57  & -0.38  & 0.98   & 0.94   & 0.99   & 0.92   & 0.84   \\
 BPT$\alpha$  &  K03 & log \oiii/\Ha  & log \nii/\Ha  &  0.68  &  0.49  & +0.03  & 0.99   & 0.98   & 0.98   & 0.98   & 0.93   \\
 BPT$\alpha$  &  S06 & log \oiii/\Ha  & log \nii/\Ha  &  0.46  &  0.29  & +0.22  & 0.98   & 0.98   & 0.98   & 0.99   & 0.93   \\
\hline     
 BPTo2        &  K01 & log \oiii/\oii & log \nii/\Ha  &  1.25  &  0.48  & -0.21  & 0.93   & 0.91   & 0.98   & 0.91   & 0.80   \\
 BPTo2        &  K03 & log \oiii/\oii & log \nii/\Ha  &  1.10  &  0.33  & +0.11  & 0.97   & 0.93   & 0.96   & 0.96   & 0.84   \\
 BPTo2        &  S06 & log \oiii/\oii & log \nii/\Ha  &  1.06  &  0.26  & +0.24  & 0.97   & 0.94   & 0.93   & 0.98   & 0.83   \\  \hline 
\hline     
 EW$\alpha$n2 &  K01 &  --            & log \nii/\Ha  & -0.10  &  --    &   --   & 0.98   & 0.82   & 0.97   & 0.87   & 0.67   \\
 EW$\alpha$n2 &  K03 &  --            & log \nii/\Ha  & -0.32  &  --    &   --   & 0.97   & 0.91   & 0.95   & 0.95   & 0.79   \\
 EW$\alpha$n2 &  S06 &  --            & log \nii/\Ha  & -0.40  &  --    &   --   & 0.96   & 0.93   & 0.92   & 0.97   & 0.80   \\
\hline
\end{tabular}
\label{tab:DD-SF-AGN} 
\end{table*}

We now start the transposition of classification criteria from
BPT-based fiducial SF/AGN classification schemes.  Three SF/AGN border
lines are considered: K01, K03 and S06. All of these can be cast onto
a single parametric form:

\begin{equation}
\label{eq:SF_x_AGN_formula}
y = a + \frac{b}{c + x}
\end{equation}

\ni where $x \equiv \log \nii/\Ha$ and $y \equiv \log
\oiii/\Hb$. These lines are drawn in Fig.\ \ref{fig:WLGsOnBPT}, and
the values of $a$, $b$, and $c$ are listed in Table
\ref{tab:DD-SF-AGN}\footnote{For S06, equation
\ref{eq:SF_x_AGN_formula} is actually a reparametrization of their
equation 11.}.  We then consider other diagnostic diagrams, and
seek a dividing line $y = f(x)$ which best maps the pre-defined
BPT-based classification scheme onto the $y \times x$ plane of the new
diagram.

For example, to separate left and right wing galaxies in the
BPT$\alpha$ diagram (Fig. \ref{fig:WLGsOnBPTa}) we use equation
\ref{eq:SF_x_AGN_formula} with $y \equiv \log \oiii/\Ha$ and $x \equiv
\log \nii/\Ha$, tagging points below and above this line as SF and
AGN, respectively, and search for values of $a$, $b$ and $c$ which
best reproduce the SF/AGN classification scheme of, say, S06.  This
optimization is achieved by identifying the coefficients $a$, $b$ and
$c$ which maximize the product (${\cal P}$) of completeness (${\cal
C}$) and reliability (${\cal R}$) fractions:

\begin{equation}
{\cal P} = {\cal C}_{\rm SF} {\cal R}_{\rm SF} {\cal C}_{\rm AGN} 
{\cal R}_{\rm AGN}
\end{equation}

\noindent where ${\cal C}_{\rm SF}$ is the fraction of galaxies
classified as SF according to the original line, say, S06, which are
correctly classified as such by our new dividing line, and ${\cal
R}_{\rm SF}$ is the fraction of objects below the new $y=f(x)$ line
which are also S06-SF, while ${\cal C}_{\rm AGN}$ and ${\cal R}_{\rm
AGN}$ are the corresponding completeness and reliability fractions for
AGN.  Except for strong covariances among $a$, $b$ and $c$, which
require some numerical care but are irrelevant for the purposes of
identifying an effective dividing line, this is a straightforward
procedure to translate a well established classification criterion to
an alternative one.

The same procedure is applied to the BPTo2 diagram. The results are
summarized in Table ~\ref{tab:DD-SF-AGN}, which lists the meaning of
$y$ and $x$, and the values of $a$, $b$ and $c$ for the BPT$\alpha$
and BPTo2 diagrams, for which SF/AGN dividing lines in the form of
equation \ref{eq:SF_x_AGN_formula} are suitable.  The resulting
transformations are very efficient, with completeness (${\cal C}$) and
reliability (${\cal R}$) factors ranging from 91 to 99\% (Table
\ref{tab:DD-SF-AGN}).

\begin{figure}
\includegraphics[bb= 30 170 250 690,width=0.5\textwidth]{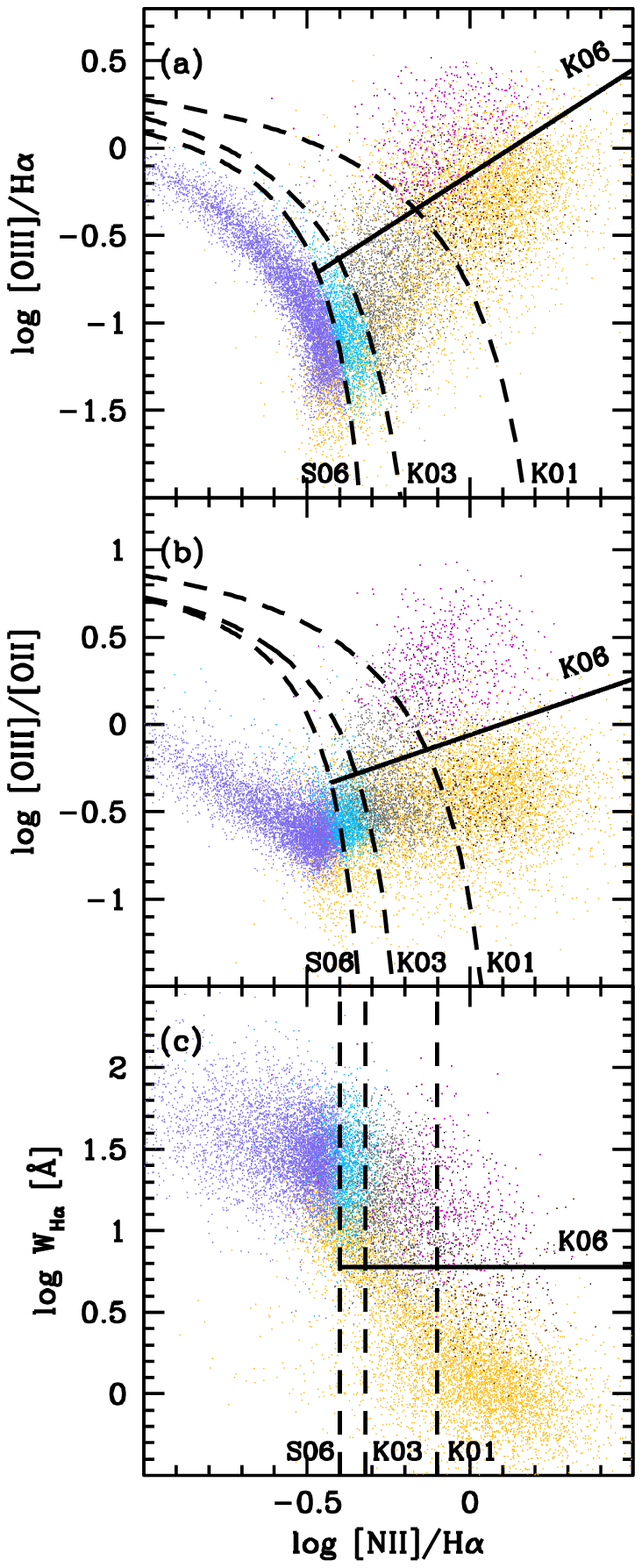}
\caption{BPT$\alpha$ (a), BPTo2 (b) and EW$\alpha$n2 (c) diagnostic
diagrams, showing the transposed SF/AGN border lines of S06, K03 and
K01 in dashed lines. Solid lines show the transposed Seyfert/LINER
classification of K06 (see Section \ref{sec:SummaryTransposedLines}
for the corresponding equations). SLGs are plotted following the same
color-coding of Fig.\ \ref{fig:WLGsOnBPT}a.  WLGs are plotted in
orange.}
\label{fig:OurDDsAndDividingLines}
\end{figure}

Figs.\ \ref{fig:OurDDsAndDividingLines}a and b show as dashed lines
the transposed S06, K03 and K01 SF/AGN boundaries in the BPT$\alpha$
and BPTo2 diagrams.  In both diagrams the transposed S06 and K03 lines
yield slightly better results (higher ${\cal P}$) than K01. This
happens for the already noted reason that the S06 and K03 lines
intercept the data at more vertical angles than the K01 one, being
therefore less prone to reddening effects, which act vertically in
both the BPT$\alpha$ and BPTo2 diagrams.

\subsection{Parameters for Seyfert/LINER borderlines in the BPT, BPT$\alpha$ and BPTo2 diagrams}

\label{sec:Sey_LINER_div_lines}


\begin{table*} 
\caption{Optimal $y = a x + b$ Seyfert/LINER dividing lines}
\begin{tabular}{llclccccccc}       
\hline\hline 
 diagram  &  $y$ & $x$ & $a$ & $b$  &  ${\cal C}_{\rm L}$ & ${\cal C}_{\rm S}$ & ${\cal R}_{\rm L}$ & ${\cal R}_{\rm S}$ & ${\cal P}$ \\
\hline 
 BPT          & log \oiii/\Hb  &  log \nii/\Ha   & 1.01  &  0.48  & 0.95   & 0.92   & 0.95   & 0.93   & 0.77   \\
 BPT$\alpha$  & log \oiii/\Ha  &  log \nii/\Ha   & 1.20  & -0.15  & 0.91   & 0.85   & 0.90   & 0.85   & 0.59   \\
 BPTo2        & log \oiii/\oii &  log \nii/\Ha   & 0.64  & -0.06  & 0.98   & 0.96   & 0.98   & 0.97   & 0.88   \\
 EW$\alpha$n2 &  $W_{\Ha}$ &  --             & --    &  6 \AA & 0.91   & 0.85   & 0.90   & 0.86   & 0.60   \\
\hline
\end{tabular}
\label{tab:DD-Seyfert-LINER} 
\end{table*}

K06 performed a detailed study of right wing sources in the SDSS,
which lead them to propose a new set of criteria to tell Seyferts from
LINERs.  These criteria are based on an empirical mapping of the
bimodality observed in the \oiii/\Hb versus \oi/\Ha and \sii/\Ha
diagrams for objects with $SN_\lambda \ge 3$ (also visible in the BPT,
but less clearly so), where AGN bifurcate into Seyfert and LINER
branches.  After filtering out SF and putative SF$+$AGN composites by
requiring objects to lie above the K01 extreme starburst lines in the
BPT, \oi/\Ha and \sii/\Ha diagrams, they define Seyferts and LINERs
using linear division lines in the $\log \oiii/\Hb $ versus $ \log
\oi/\Ha$ and $\log \oiii/\Hb $ versus $ \log \sii/\Ha$ spaces.

This scheme has a couple of caveats. First, as is common with
classification schemes involving more than one diagnostic diagram,
inconsistencies abound. In this particular case, galaxies with
ambiguous classification are as numerous as properly classified
Seyferts and LINERs. A second drawback of the K06 scheme is that, as
seen in Fig.\ \ref{fig:S2N_distribs}, it requires far more good
quality emission line data than one can usually afford with SDSS-like
spectra.

The technique explained above offers an opportunity to remedy this
situation, translating the K06 classification scheme into simpler,
more economic and thus more widely applicable criteria.

We start deriving a Seyfert/LINER classification criterion based {\em
exclusively on the BPT diagram}. We compute the values of the
coefficients of a simple straight line in the BPT plane which maximize
the product ${\cal P} = {\cal C}_{\rm L} {\cal R}_{\rm L} {\cal
C}_{\rm S} {\cal R}_{\rm S}$, where ${\cal C}_{\rm S}$ (${\cal C}_{\rm
L}$) is the fraction of K06-Seyfert (LINER) galaxies which are
correctly classified as such by our dividing line, and ${\cal R}_{\rm
S}$ (${\cal R}_{\rm L}$) is the fraction of objects above (below) our
line which are also Seyfert (LINER) according to K06.  Sources with an
ambiguous classification are not used in this transposition.  We find
that a border line

\begin{equation}
\label{eq:Seyfert_LINER_formula_BPT}
\log \frac{\oiii}{\Hb} = 1.01 \log \frac{\nii}{\Ha} + 0.48
\end{equation}

\ni does a good job in translating the K06 Seyfert/LINER
classification to the BPT diagram, with completeness and reliability
fractions of 92\% or better (Table \ref{tab:DD-Seyfert-LINER}). This
line is drawn in Fig.\ref{fig:WLGsOnBPT}. Our derived parameters are
very similar to the $a = 1.05$ and $b = 0.45$ adopted by
\citet{Schawinski_etal_2007}.

The same transposition procedure was carried out for the BPT$\alpha$
and BPTo2 diagrams, with results shown as solid lines in the top two
panels of Fig.\ \ref{fig:OurDDsAndDividingLines}.  As for the BPT
diagram, a simple $y = ax + b$ border line suffices to separate
Seyferts from LINERs in these diagrams. 

Table ~\ref{tab:DD-Seyfert-LINER} lists the optimal values of $a$ and
$b$ for the BPT, BPT$\alpha$ and BPTo2 diagrams.  As in the case of
the SF/AGN classification, the quality of these transformations can be
measured by the completeness and reliability factors, which range from
85 to 98\% (Table \ref{tab:DD-Seyfert-LINER}).  The best results are
achieved for the BPTo2 diagram (${\cal P} = 0.88$), followed by the
BPT ($0.77$) and BPT$\alpha$ ($0.59$), corroborating the qualitative
assessment of the Seyfert/LINER diagnostic power of these diagrams
presented in Section \ref{sec:DDforWLGs}.

The addition of WLGs to the BPT and BPT$\alpha$ diagrams causes
significant dilution of the Seyfert/LINER dichotomy, leading to the
nagging suspicion that selection effects might be behind what is
presumed to be a physical class-separation.  The fact that the
bimodality is present in the BPTo2 plane (Fig.\
\ref{fig:OurDDsAndDividingLines}b) should dismiss such worries. The
inclusion of WLGs, however, does lead to a new perspective on the
Seyfert/LINER bimodality, as discussed in Section
\ref{sec:K06Bimodality}.

\subsection{Comments on combinations of Seyfert/LINER and SF/AGN criteria}

Accepting that AGN come in either Seyfert or LINER flavours, one is
lead to a classification scheme based on three fundamental classes: SF
galaxies, Seyferts and LINERs. Complementing the S06 SF/AGN
classification with the Seyfert/LINER division schemes derived above
thus leads to S06-SF, S06-Seyferts and S06-LINERs, and similarly for
K03 and K01.

A caveat with these combinations is that our Seyfert/LINER dividing
lines were calibrated exclusively on the basis of the K06 criteria,
which in turn were defined only for K01-AGN, and thus comprise but a
sub-set of S06 and K03-AGN.  We have expressed strong reservations
with regard to the K01 SF/AGN scheme (Section
\ref{sec:MeaningOfDivLines}), but these reservations do {\it not}
extend to the K06 Seyfert/LINER classification scheme, which is rooted
on empirical evidence of a bimodality in emission line diagnostic
diagrams for SLGs.  Nonetheless, rigorously speaking, the K06
Seyfert/LINER division applies only to K01-AGN, and extending it to
S06 and K03 involves an {\em extrapolation} to intermediate zones in
diagnostic diagrams, where the bimodality becomes fuzzier.

As can be seen in the BPT (Fig.\ \ref{fig:WLGsOnBPT}), BPT$\alpha$
(Fig.\ \ref{fig:OurDDsAndDividingLines}a) and BPTo2, (Fig.\
\ref{fig:OurDDsAndDividingLines}b), the extrapolated Seyfert/LINER
demarcation lines do not cut the right wing in equal halves. Sources
on the LINER side of these lines become more common considering the
whole population than when restricting to galaxies above the K01 line.

\subsection{Dividing lines in the EW$\alpha$n2 diagram}

\begin{figure}
\includegraphics[bb= 30 170 350 690,width=0.5\textwidth]{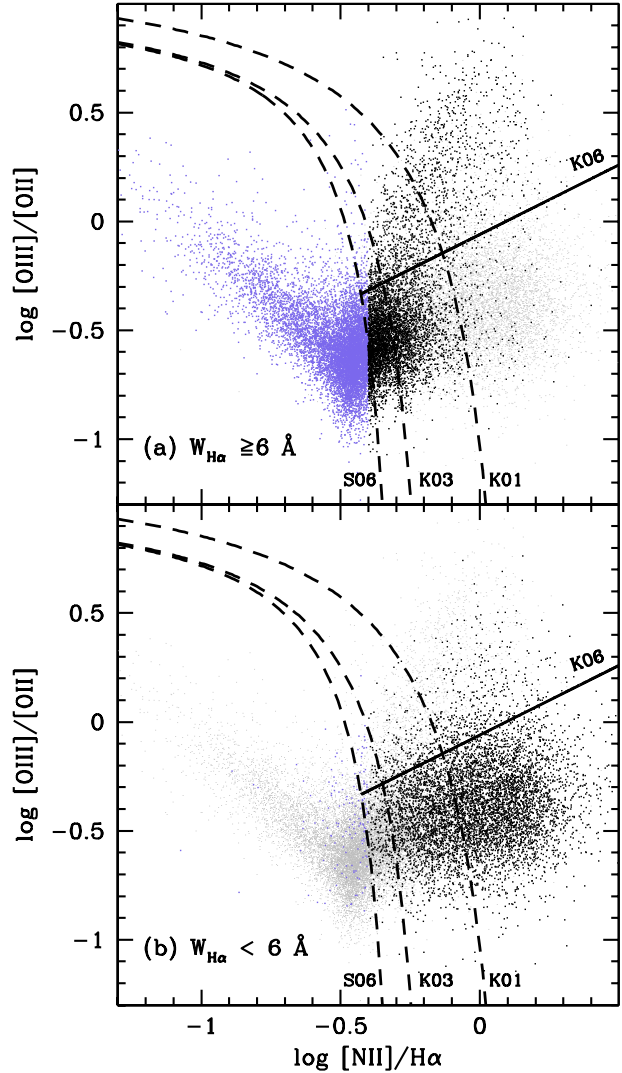}
\caption{BPTo2 diagram indicating with black points the location of
galaxies below (a) and above (b) the optimal separation Seyfert/LINER
separation in terms of the equivalent width of \Ha: $W_{\Ha} = 6.0$
\AA. Points in violet and black mark S06-SF and S06-AGN according to
the $\log \nii/Ha < $ and $\ge -0.40$ criteria, respectively. For
reference, the background points (light grey) show all galaxies,
irrespective of $W_{\Ha}$.  Only $SN_\lambda \ge 3$ data (in \oii,
\oiii, \Ha and \nii) are used in these plots.  The dividing lines are
the same as in Fig.\ \ref{fig:OurDDsAndDividingLines}b (equations
\ref{eq:TranspDivLine_S06_BPTo2}, \ref{eq:TranspDivLine_K03_BPTo2},
\ref{eq:TranspDivLine_K01_BPTo2}, and
\ref{eq:TranspDivLine_K06_BPTo2}).
}
\label{fig:DDs_X_EWHaEQ6Cuts}
\end{figure}

A simpler transposition strategy was used to deal with the
EW$\alpha$n2 diagram. As discussed in Section \ref{sec:ewan2} and
shown in Fig.\ \ref{fig:WLGsOnEWhan2}, the SF/AGN diagnostic power in
this case lies almost exclusively on the horizontal axis, while the
differentiation of Seyferts and LINERs occurs in the vertical
direction.  Fitting $y(x)$ dividing lines to this diagram would thus
be an exaggeration, given this one-dimensional behaviour.  Apart from
this simplification, the same optimal separator technique described
above was used to identify class separation boundaries.

The division of SFs and AGN according to the S06 BPT-based scheme is
best transposed to $\log \nii/\Ha = -0.40$, while the optimal boundary
for the K03 division is $\log \nii/\Ha = -0.32$. As anticipated
(Fig.~\ref{fig:WLGsOnEWhan2}), the least satisfactory results are
obtained for the K01 line, whose formally best division at $\log
\nii/\Ha = -0.10$ miss-classifies about 18\% of the Seyferts.

The separation of Seyfert and LINER classes as defined by K06 is best
accomplished setting a boundary at $W_{\Ha} = 6.0$ \AA, in agreement
with the histograms in Fig.\ \ref{fig:EWha_distribs}. The completeness
and reliability fractions are marginally better than for the
BPT$\alpha$, but smaller than for the BPT and BPTo2 diagrams (Table
\ref{tab:DD-Seyfert-LINER}).

Overall, compared to standard diagnostic diagrams, the EW$\alpha$n2
diagram offers an attractive compromise between classification
efficiency and economical aspects.

Fig.\ \ref{fig:DDs_X_EWHaEQ6Cuts} shows the location of galaxies below
(panel a) and above (panel b) the $W_{\Ha} = 6$ \AA\ threshold in our
alternative BPTo2 diagram for sources where \oii, \oiii, \Ha and \nii
are all detected with $SN_\lambda \ge 3$.  The plot shows that most
sources in the K01-LINERs region of this diagram, ie., those in the
bottom-right ``corner'' below the K06 line and to the right of the K01
line, indeed have $W_{\Ha} < 6$ \AA\ (points in black in the panel b,
and in light grey in panel a). Similarly, the zone corresponding to
K01-Seyferts is populated predominantly by galaxies with $W_{\Ha} > 6$
\AA.  This consistency is expected, since the K06 criteria were used
to calibrate the border lines in both diagrams.  Notice that $W_{\Ha}
< 6$ \AA\ also eliminates nearly all S06 and K03 SF systems, even
though this criterion was not explicitly designed to do so. In the
intermediate zone between the K01 line and the S06 lines, sources with
$W_{\Ha}$ below and above the 6 \AA\ cut become heavily mixed. The
extrapolation of the K06-based division line in the BPTo2 plane places
most such intermediate sources in the LINER branch, while the
$W_{\Ha}$ cut suggest a more even mix of Seyferts and LINERs.

\subsection{Summary of SF/AGN and Seyfert/LINER dividing lines}
\label{sec:SummaryTransposedLines}

For ease of use, we open up the results summarized in Tables and
\ref{tab:DD-SF-AGN} and \ref{tab:DD-Seyfert-LINER} into to explicit
equations for emission line classification (shown in Fig.\
\ref{fig:OurDDsAndDividingLines}).

The S06, K03 and K01 SF/AGN dividing lines in the BPT plane are best
transposed to the BPT$\alpha$ and BPTo2 diagrams through

\begin{equation}
\label{eq:TranspDivLine_S06_BPTa}
\log \frac{\oiii}{\Ha} = 0.46 + \frac{0.29}{\log \frac{\nii}{\Ha} + 0.22}
\hfill {\rm (BPT\alpha-S06)~~~~}
\end{equation}

\begin{equation}
\label{eq:TranspDivLine_S06_BPTo2}
\log \frac{\oiii}{\oii} = 1.06 + \frac{0.26}{\log \frac{\nii}{\Ha} + 0.24}
\hfill {\rm (BPTo2-S06)~~~~}
\end{equation}

\begin{equation}
\label{eq:TranspDivLine_K03_BPTa}
\log \frac{\oiii}{\Ha} = 0.68 + \frac{0.49}{\log \frac{\nii}{\Ha} + 0.33}
\hfill {\rm (BPT\alpha-K03)~~~~}
\end{equation}

\begin{equation}
\label{eq:TranspDivLine_K03_BPTo2}
\log \frac{\oiii}{\oii} = 1.10 + \frac{0.33}{\log \frac{\nii}{\Ha} + 0.11}
\hfill {\rm (BPTo2-K03)~~~~}
\end{equation}

\begin{equation}
\label{eq:TranspDivLine_K01_BPTa}
\log \frac{\oiii}{\Ha} = 0.69 + \frac{0.57}{\log \frac{\nii}{\Ha} - 0.38}
\hfill {\rm (BPT\alpha-K01)~~~~}
\end{equation}

\begin{equation}
\label{eq:TranspDivLine_K01_BPTo2}
\log \frac{\oiii}{\oii} = 1.25 + \frac{0.48}{\log \frac{\nii}{\Ha} - 0.21}
\hfill {\rm (BPTo2-K01)~~~~}
\end{equation}

An alternative (and cheaper) way to distinguish SF from AGN is through
the \nii/\Ha ratio.  The limiting values are $\log \nii/\Ha = -0.40$,
$-0.32$, and $-0.10$ for the S06, K03 and K01 SF/AGN schemes,
respectively. Recall that, for the reasons discussed in Section
\ref{sec:MeaningOfDivLines}, the K01 classification scheme is
misleading, and thus equations \ref{eq:TranspDivLine_K01_BPTa},
\ref{eq:TranspDivLine_K01_BPTo2}, and the $\log \nii/\Ha = -0.10$
criteria are {\em not recommended}.

The K06 Seyfert/LINER classification scheme can be recast onto
dividing lines in the BPT, BPT$\alpha$ and BPTo2 diagrams by

\begin{equation}
\label{eq:TranspDivLine_K06_BPT}
\log \frac{\oiii}{\Hb} = 1.01 \log \frac{\nii}{\Ha} + 0.48
\hfill {\rm (BPT)~~~~~~}
\end{equation}

\begin{equation}
\label{eq:TranspDivLine_K06_BPTa}
\log \frac{\oiii}{\Ha} = 1.20 \log \frac{\nii}{\Ha} - 0.15
\hfill {\rm (BPT\alpha)~~~~~~}
\end{equation}

\begin{equation}
\label{eq:TranspDivLine_K06_BPTo2}
\log \frac{\oiii}{\oii} = 0.64 \log \frac{\nii}{\Ha} - 0.06
\hfill {\rm (BPTo2)~~~~~~}
\end{equation}

\ni of which the last one (BPTo2) is the most recommended.

In the absence of reliable \oiii and \oii fluxes, $W_{\Ha}$ does an
acceptable job in distinguishing Seyferts from LINERs, with an optimal
border line located at $W_{\Ha} = 6$ \AA.



\section{The ELG population in light of our alternative classification}
\label{sec:Application}

With the alternative and more inclusive classification schemes
outlined above, we can finally place WLGs onto standard spectral
categories and abandon the WL-H, WL-HO and WL-O notation. This is the
topic of this section, which also discusses how the inclusion of this
population changes the balance of galaxy spectral classes in the
nearby Universe, and how WLGs affect the dichotomy between Seyferts
and LINERs.

\begin{figure*}
\centering
    \vspace{10pt}
    \includegraphics[width=.2\textwidth, angle=270, bb=100 50 530 710]{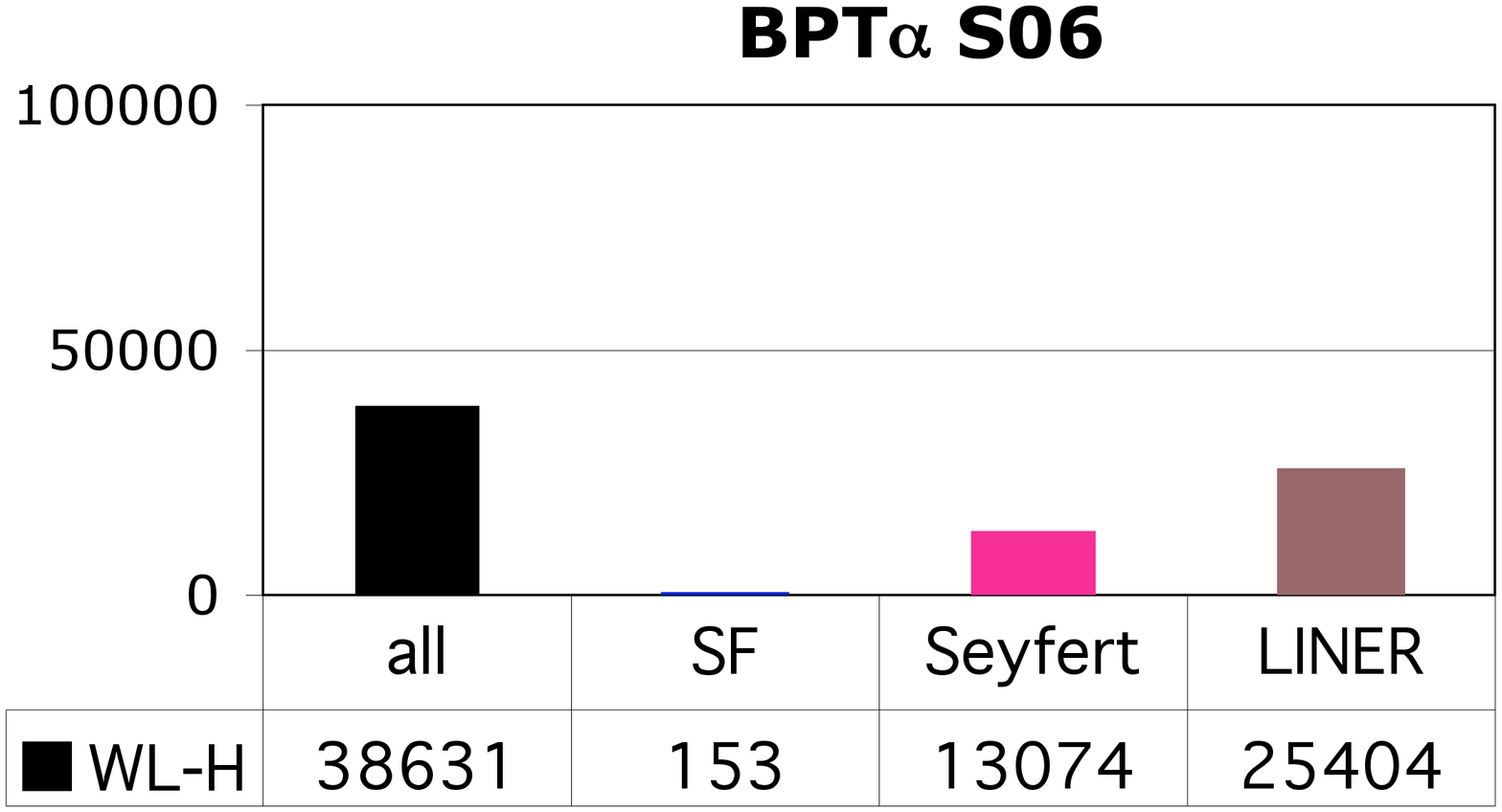}
    \includegraphics[width=.2\textwidth, angle=270, bb=100 50 530 710]{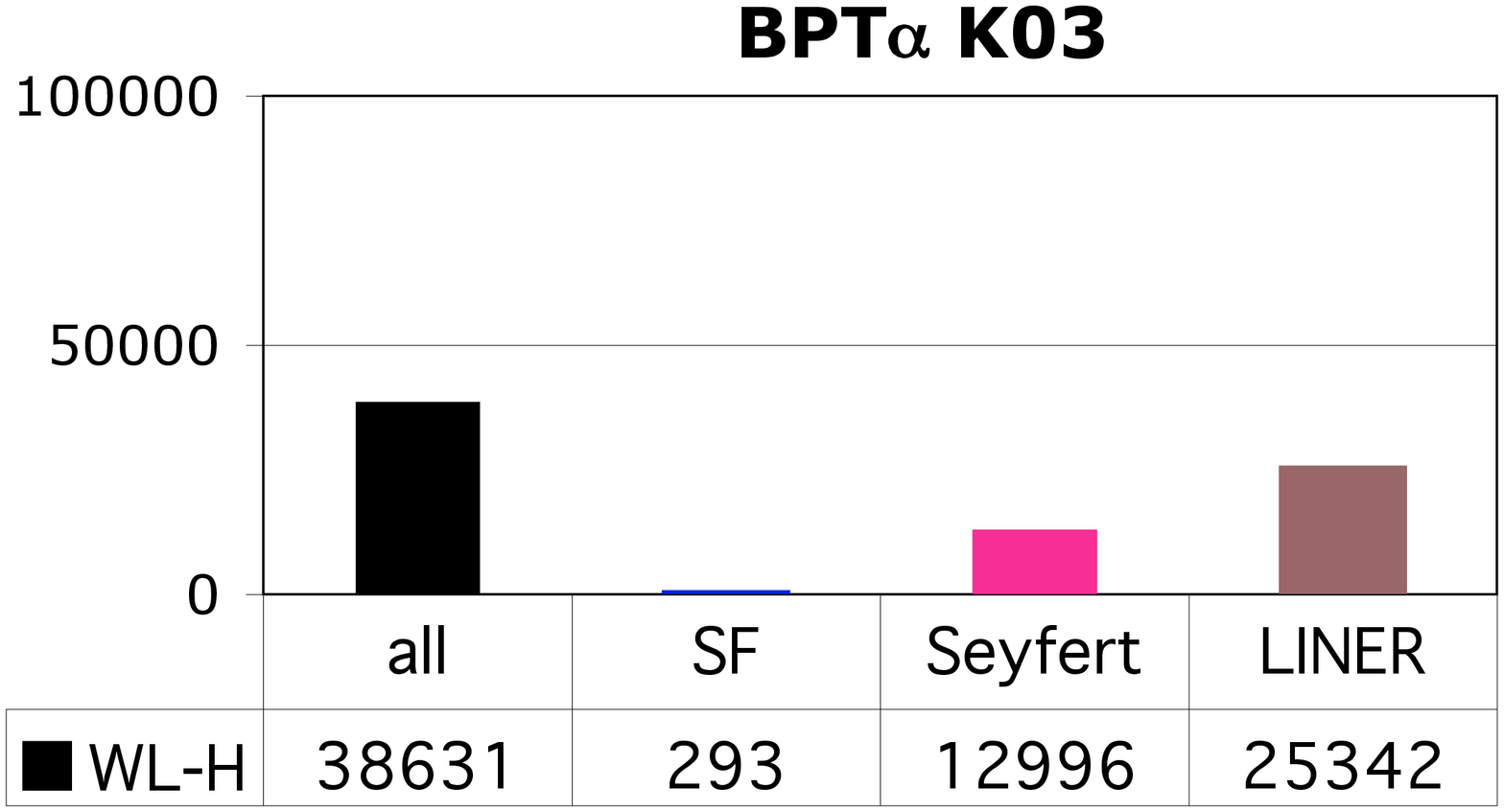}
    \includegraphics[width=.2\textwidth, angle=270, bb=100 50 530 710]{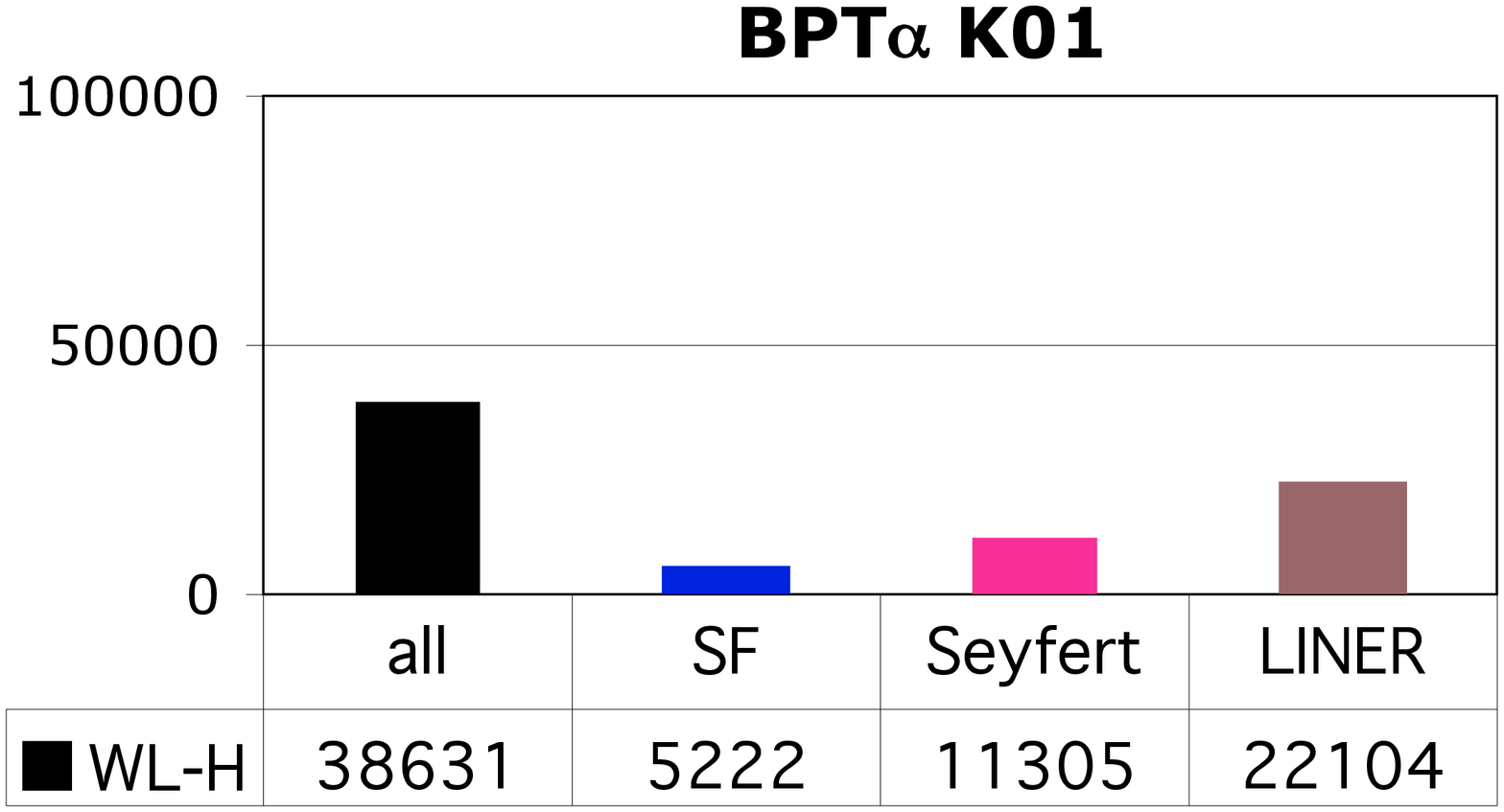}
    \vspace{10pt}
    \includegraphics[width=.2\textwidth, angle=270, bb=100 50 530 710]{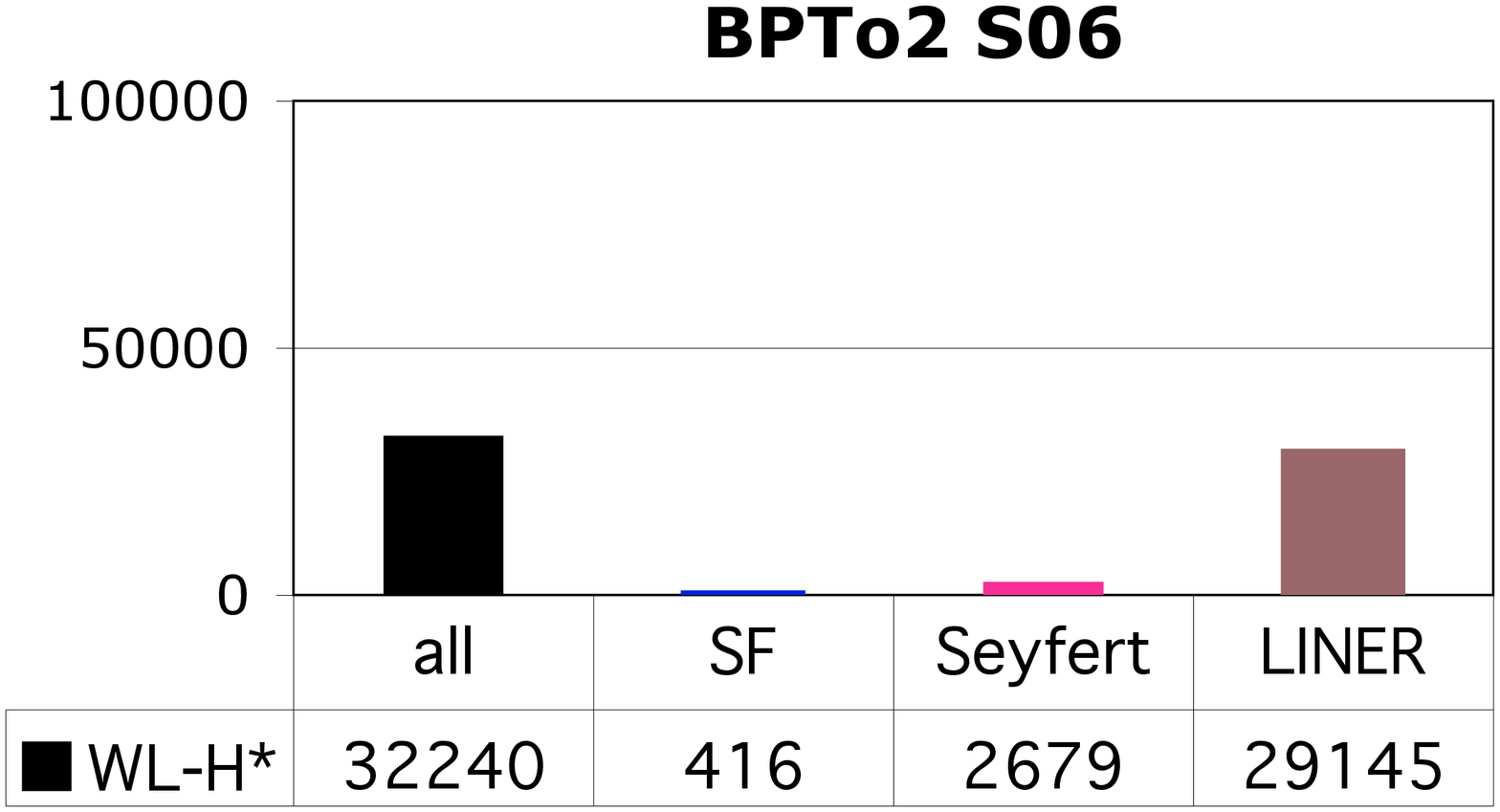}
    \includegraphics[width=.2\textwidth, angle=270, bb=100 50 530 710]{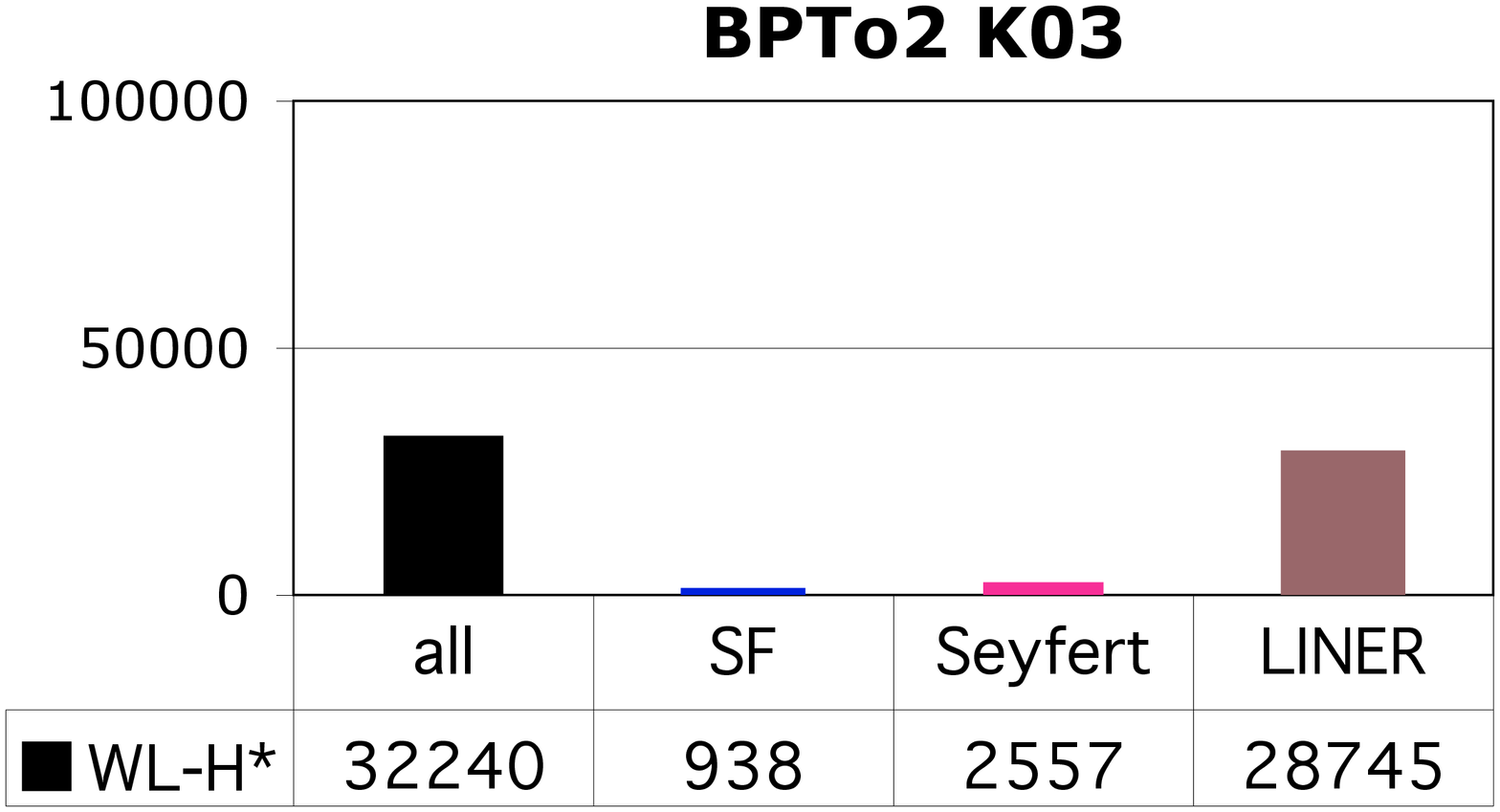}
    \includegraphics[width=.2\textwidth, angle=270, bb=100 50 530 710]{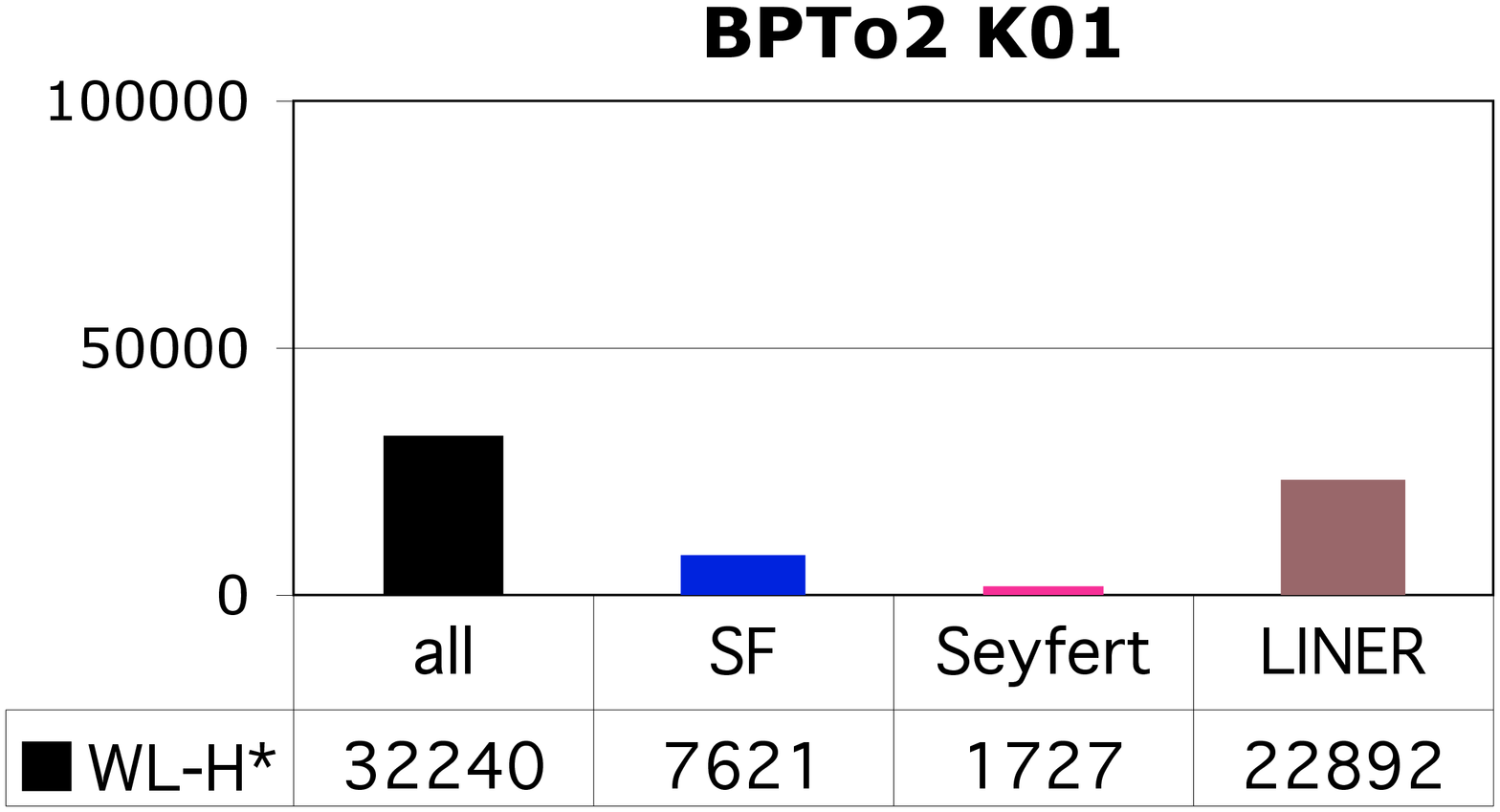}
    \includegraphics[width=.2\textwidth, angle=270, bb=100 50 530 710]{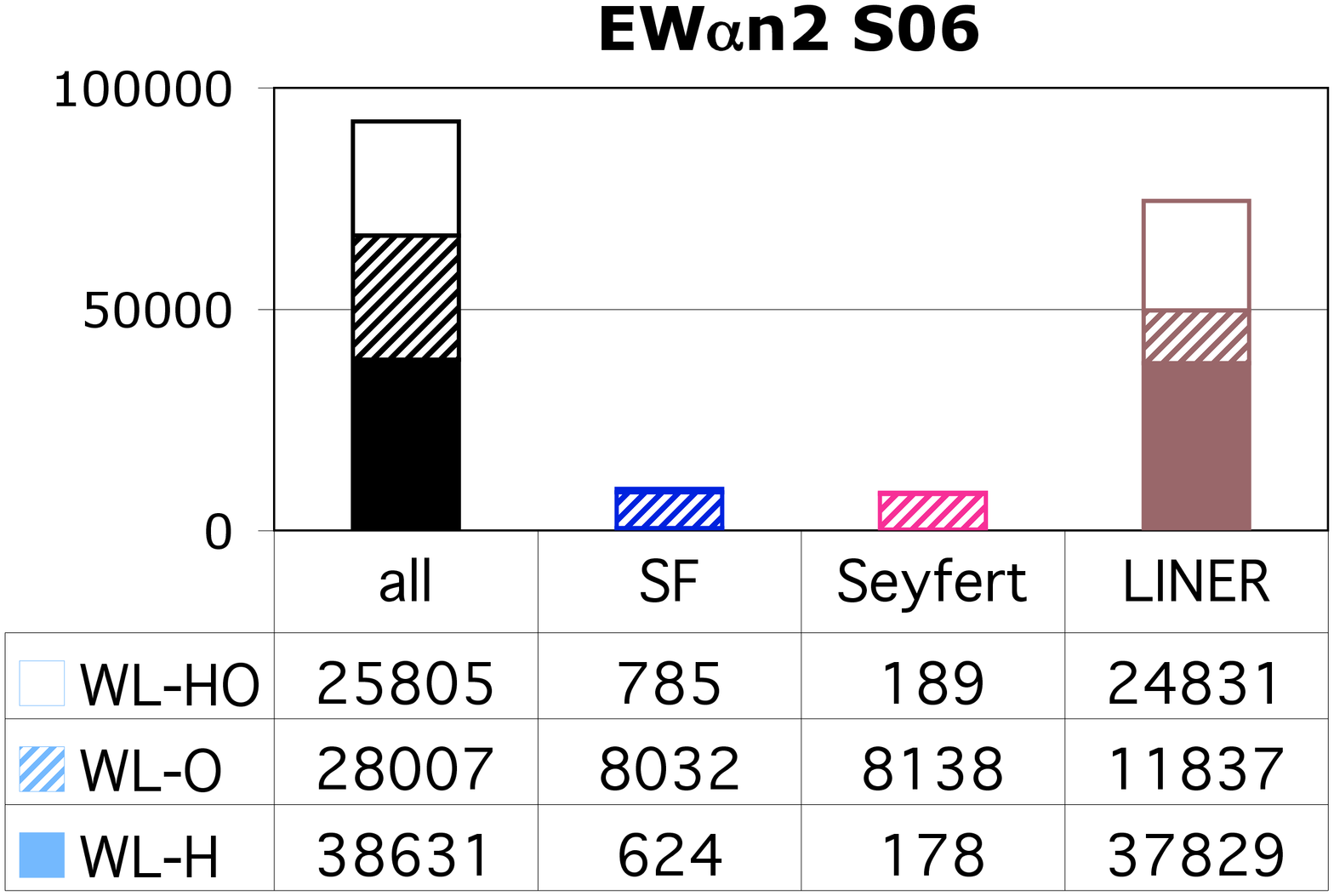}
    \includegraphics[width=.2\textwidth, angle=270, bb=100 50 530 710]{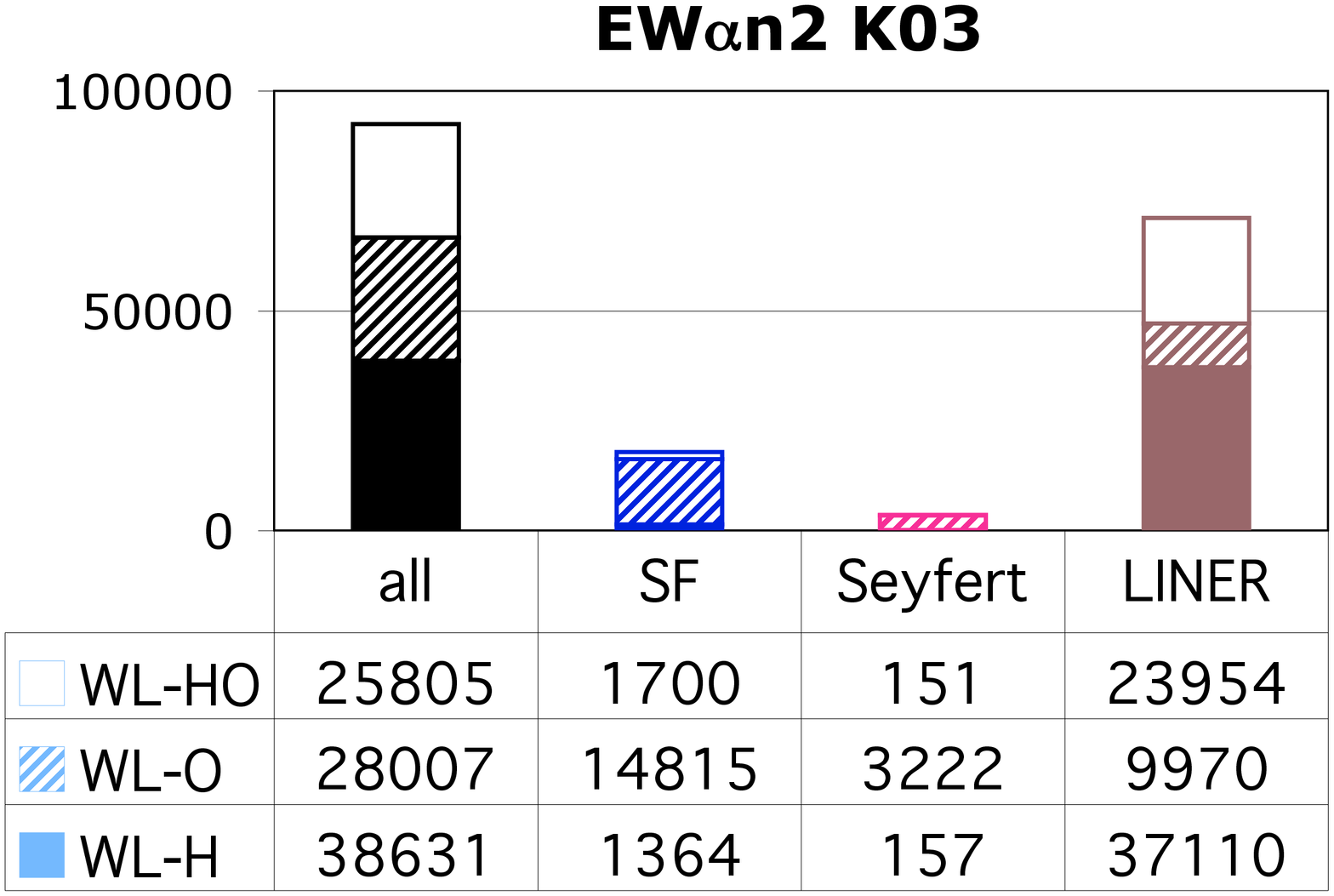}
    \includegraphics[width=.2\textwidth, angle=270, bb=100 50 530 710]{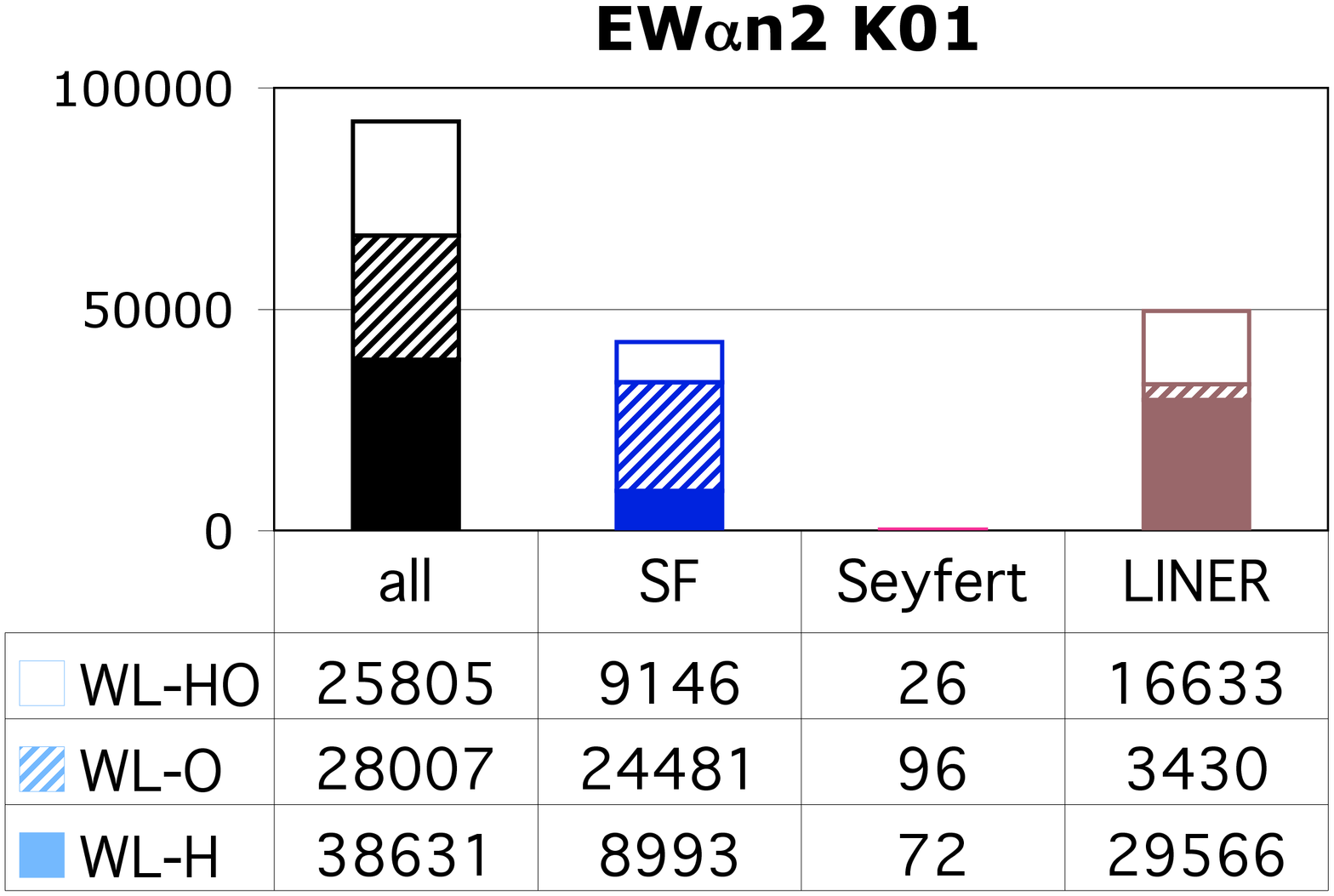}
    \vspace{10pt}
\caption{Spectral classification of WLGs according to the BPT$\alpha$
(top row), BPTo2 (middle) and EW$\alpha$n2 (bottom) diagrams. Results
for the S06, K03 and K01 classification schemes are given in the left,
middle, and right, respectively. For each diagram, only galaxies with
$SN_\lambda \ge 3$ in all lines involved are classified. In the BPTo2,
WL-H$^\star$ denotes WL-Hs which have $SN_{\oii} \ge 3$.}
\label{fig:classif_wlgs}
\end{figure*}

The three classes considered below are SF, Seyferts and LINERs. Given
our comments on the meaning of the S06, K03 and K01 divisory lines
(see Section \ref{sec:MeaningOfDivLines}), we intentionally do {\em
not} define a class of ``composite'' galaxies, even though such
systems are expected to be present in the SDSS.  The K01 scheme is not
capable of adequately confining such SF+AGN hybrids in any of the 3
classes.  {\em Ad hoc} combinations of criteria, like K01+K03, can be
postulated to select sources with intermediate line ratios, but these
are not necessarily true mixtures of SF and AGN-powered emission line
systems. The situation is less confusing with the S06 scheme, where
the SF class is designed to isolate ``pure SF'' systems, and hence
SF+AGN hybrids are confined to S06-Seyferts or S06-LINERs. The same
applies to the K03 scheme, which, according to the hybrid
photoionization models by S06, admits AGN contributions to the
ionizing power of at most 3\%.  In both these schemes, anything that
is not a pure SF counts either as a Seyfert or a LINER. This is the
best that can be done, given that unambiguous definitions of ``pure
AGN'', SF+AGN composites and other mechanisms leading to AGN-like line
ratios are not possible on the basis of optical emission line data
alone.

\subsection{Classification of WLGs}
\label{sec:ClassificationOfWLGs}

Fig.~\ref{fig:classif_wlgs} shows the results of the classification of
WLGs into SF, Seyfert and LINER classes using the BPT$\alpha$ (top),
BPTo2 (middle) and EW$\alpha$n2 (bottom) diagrams. For the distinction
between AGN from SF galaxies, the left diagrams use the divisory lines
derived by transposing the S06 criterion, while the middle and right
panels show results obtained with the K03 and K01 criteria,
respectively.

With the BPT$\alpha$ diagram, WLG galaxies can be robustly classified
only if they belong to the WL-H family, so only those are represented
in the top panels in Fig.~\ref{fig:classif_wlgs}. As clearly seen in
this figure, most WL-Hs are LINER-like, whatever the system of
divisory lines is used (S06, K03 or K01). In the K01 system, 14\% of
WL-Hs cross the border towards the SF zone, but, as already explained,
these are definitely not pure SF galaxies. The tiny numbers of S06-SF
and K03-SF among WL-Hs indicates that almost all of them are indeed
AGN-like.

For the BPTo2 diagram (middle panels), only galaxies with $SN_\lambda
\ge 3$ in \oii, \oiii, \Ha and \nii are included to ensure a robust
classification.  Here again we see that most WL-Hs are classified as
AGN, and most of them are LINER-like. As noted before (see Fig.\
\ref{fig:WLGsOnBPTo2}), the BPTo2 diagram is much more efficient than
both the BPT and the BPT$\alpha$ diagrams in distinguishing Seyferts
from LINERs. About 90\% of WL-Hs are S06-LINERs and only a tiny
proportion are S06-Seyferts. Similar fractions apply to K03.

The bottom panels of Fig.~\ref{fig:classif_wlgs} show the results
obtained with the EW$\alpha$n2 diagram. This time, we can also
classify galaxies from the WL-O and WL-HO families, whose numbers are
represented by hatched and empty areas, respectively.  Adding these
new sources further increases the number of LINER-like galaxies in the
S06 and K03 schemes. The K01 scheme shows a different picture,
dividing the WLGs almost equally among SFs and LINERs, but, as
emphasized before (see Fig.\ \ref{fig:OurDDsAndDividingLines}), the
EW$\alpha$n2 diagram is not suitable to tell K01-SF from K01-AGN. The
weak line SF galaxies identified with the S06 and K03 schemes, on the
other hand, are pure SF galaxies. They come mostly from the WL-Os,
robustly classifiable in the EW$\alpha$n2 diagram but not in previous
ones.  About 53\% of WL-Os turn out to be SF galaxies according to the
K03-scheme, a fraction which reduces to 29\% in the more stringent S06
scheme.

\subsection{A revised census of ELGs in the local Universe}
\label{sec:RevisedCensus}

We are now able to classify a significantly larger number of galaxies
than when using the BPT diagram: from $\sim 20$ to 50\% more,
depending on the alternative diagram that is used. Does this change
our view of the population of ELGs in the local Universe?

\begin{figure*}
\centering
    \includegraphics[width=.2\textwidth, angle=270, bb=100 50 530 710]{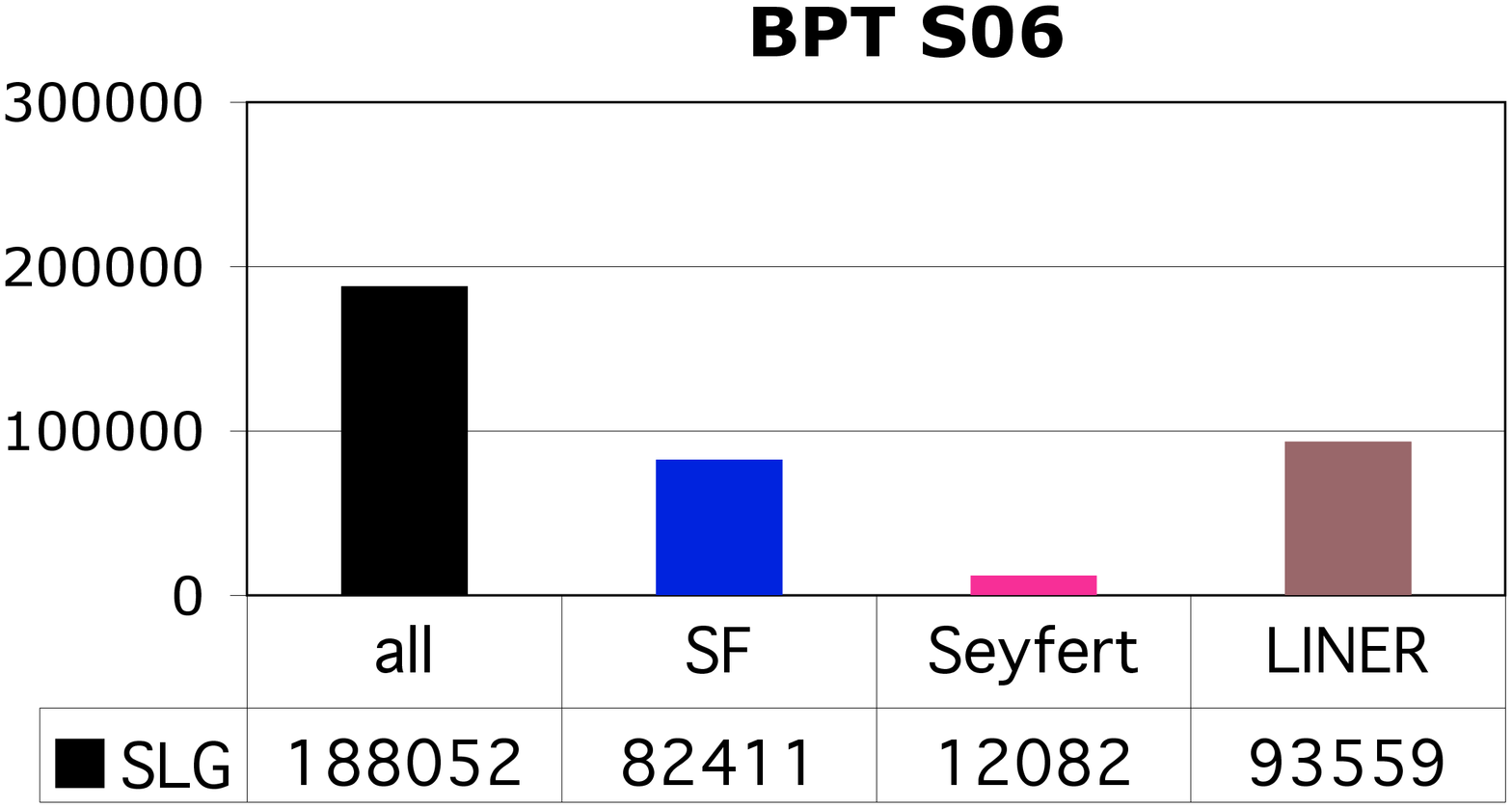}
    \includegraphics[width=.2\textwidth, angle=270, bb=100 50 530 710]{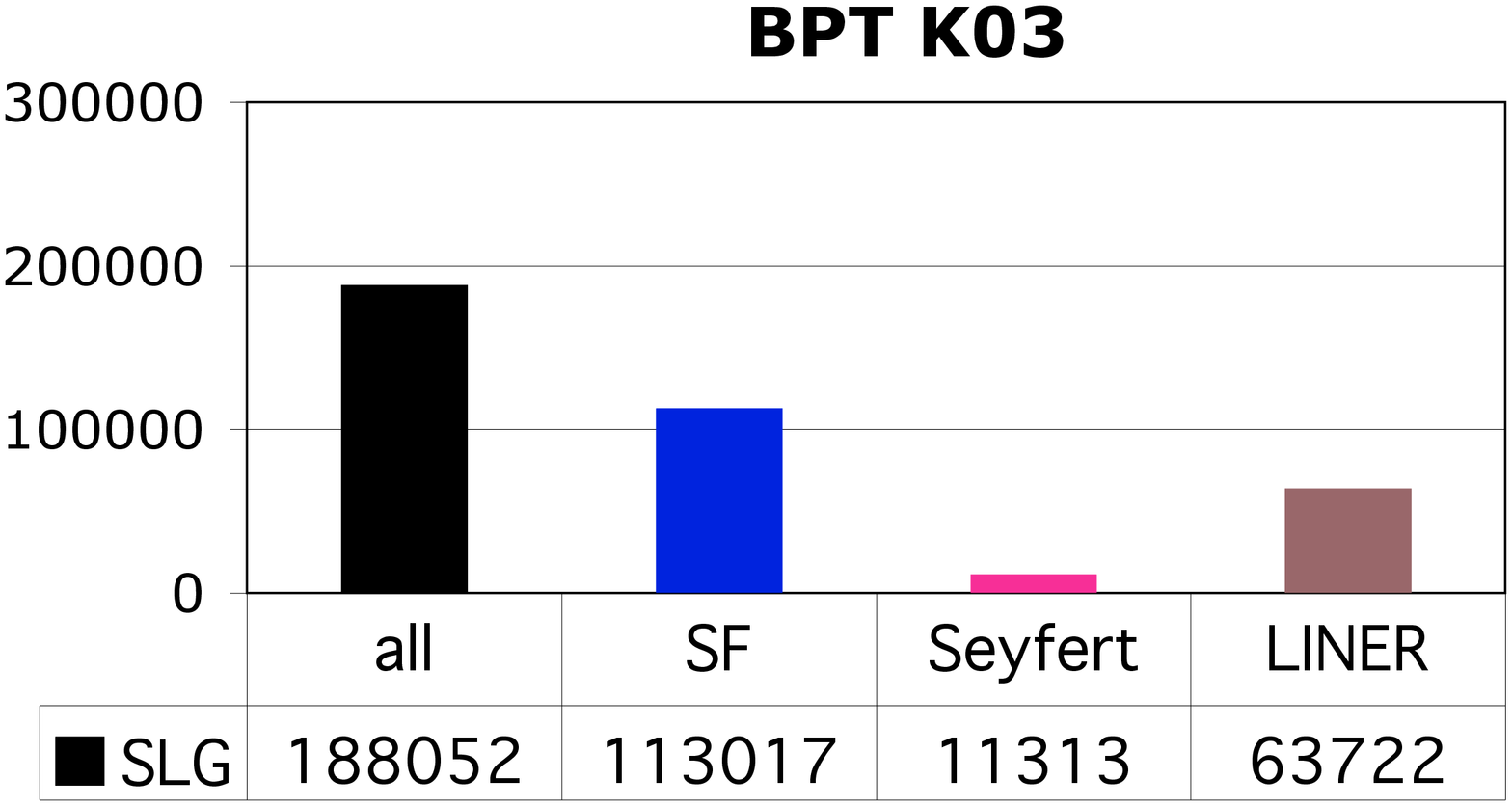}
    \includegraphics[width=.2\textwidth, angle=270, bb=100 50 530 710]{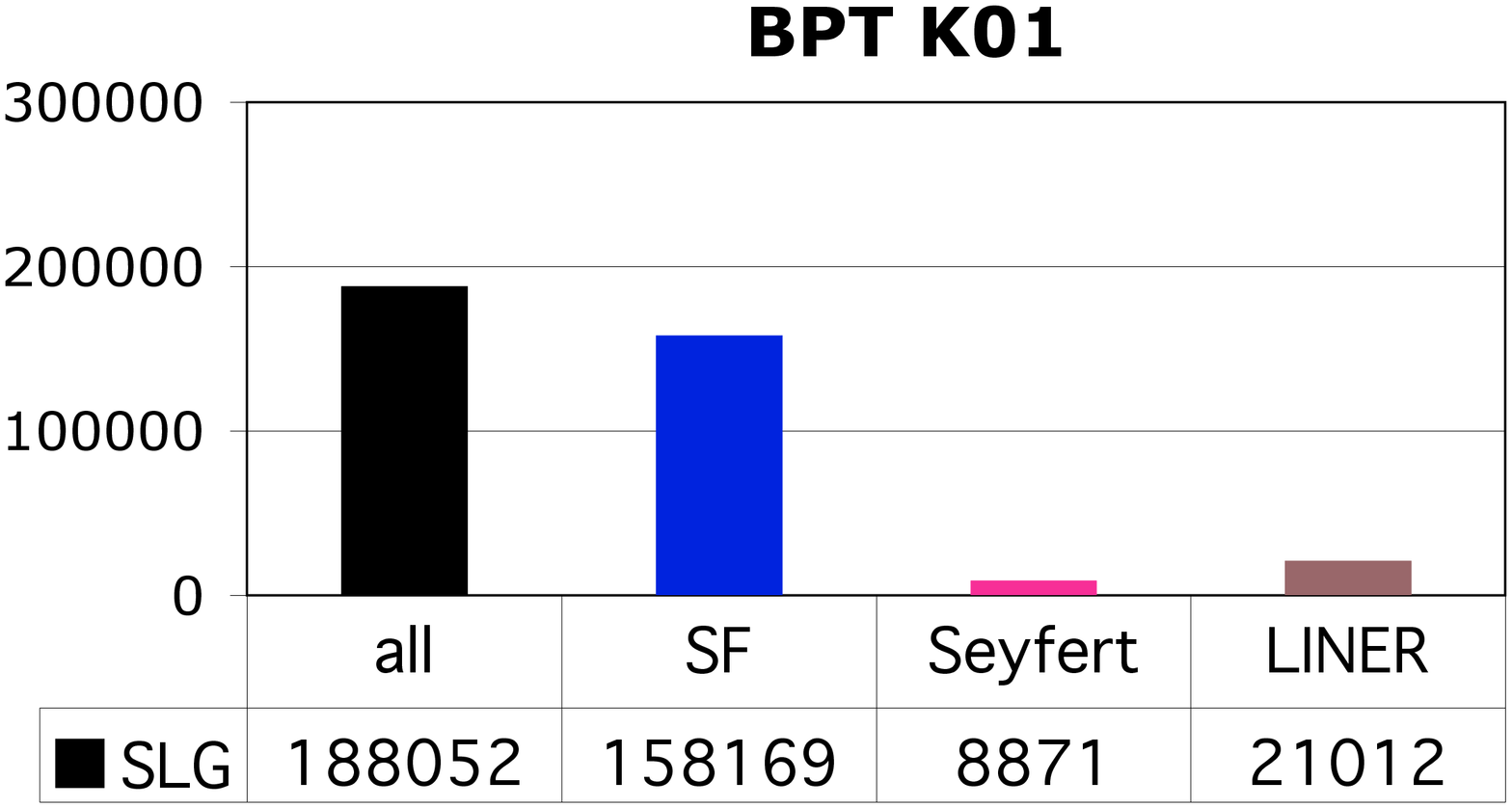}
\caption{Emission line classification according to purely BPT-based
criteria. Left, middle and right plots correspond to results obtained
with the S06, K03 and K01 criteria, respectively.  Only $SN_\lambda
\ge 3$ lines are used, which in this case correspond to our definition
of SLGs.}
\label{fig:census_O3Hb}
\end{figure*}

\begin{figure*}
\centering
    \includegraphics[width=.2\textwidth, angle=270, bb=100 50 530 710]{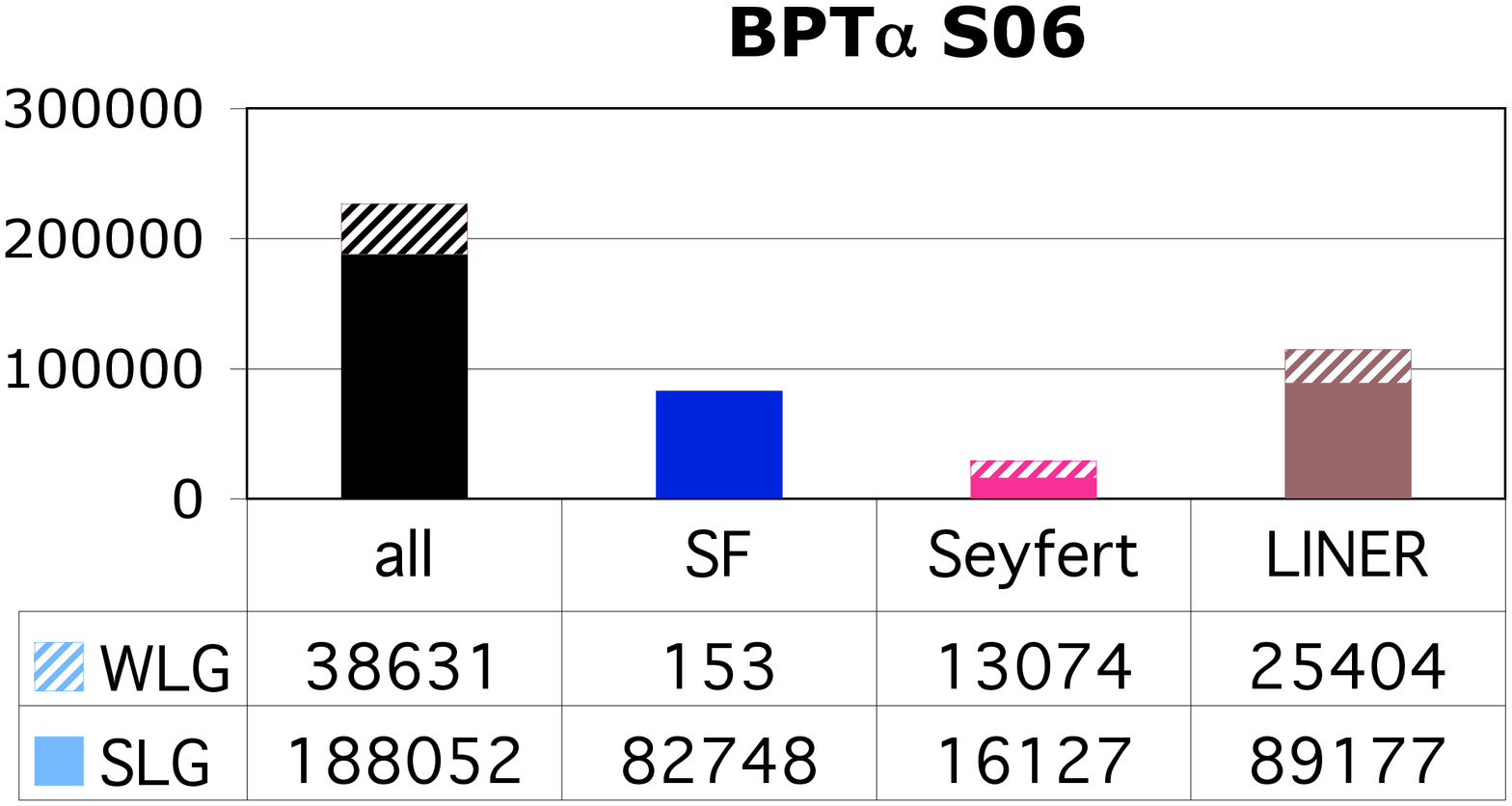}
    \includegraphics[width=.2\textwidth, angle=270, bb=100 50 530 710]{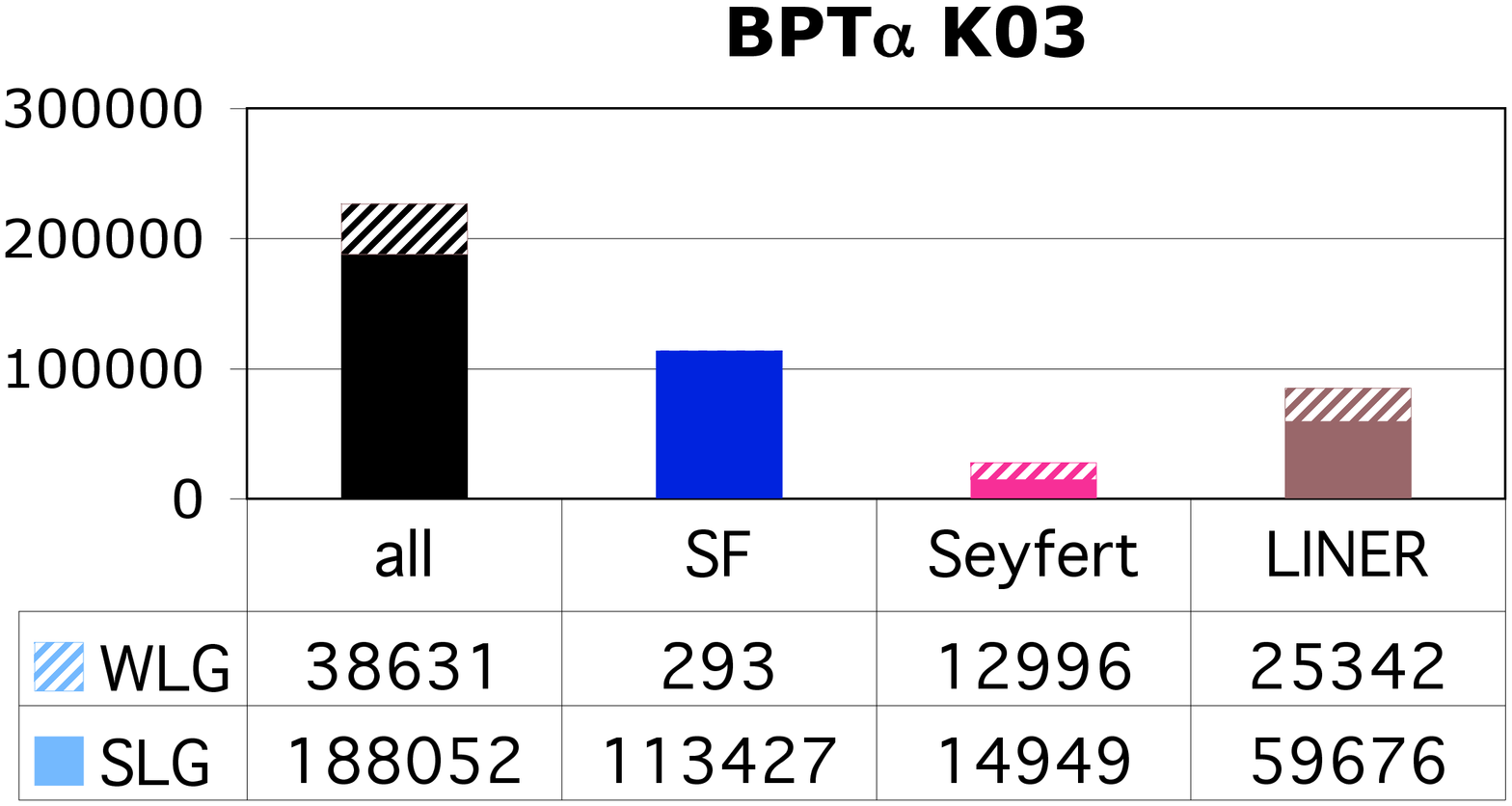}
    \includegraphics[width=.2\textwidth, angle=270, bb=100 50 530 710]{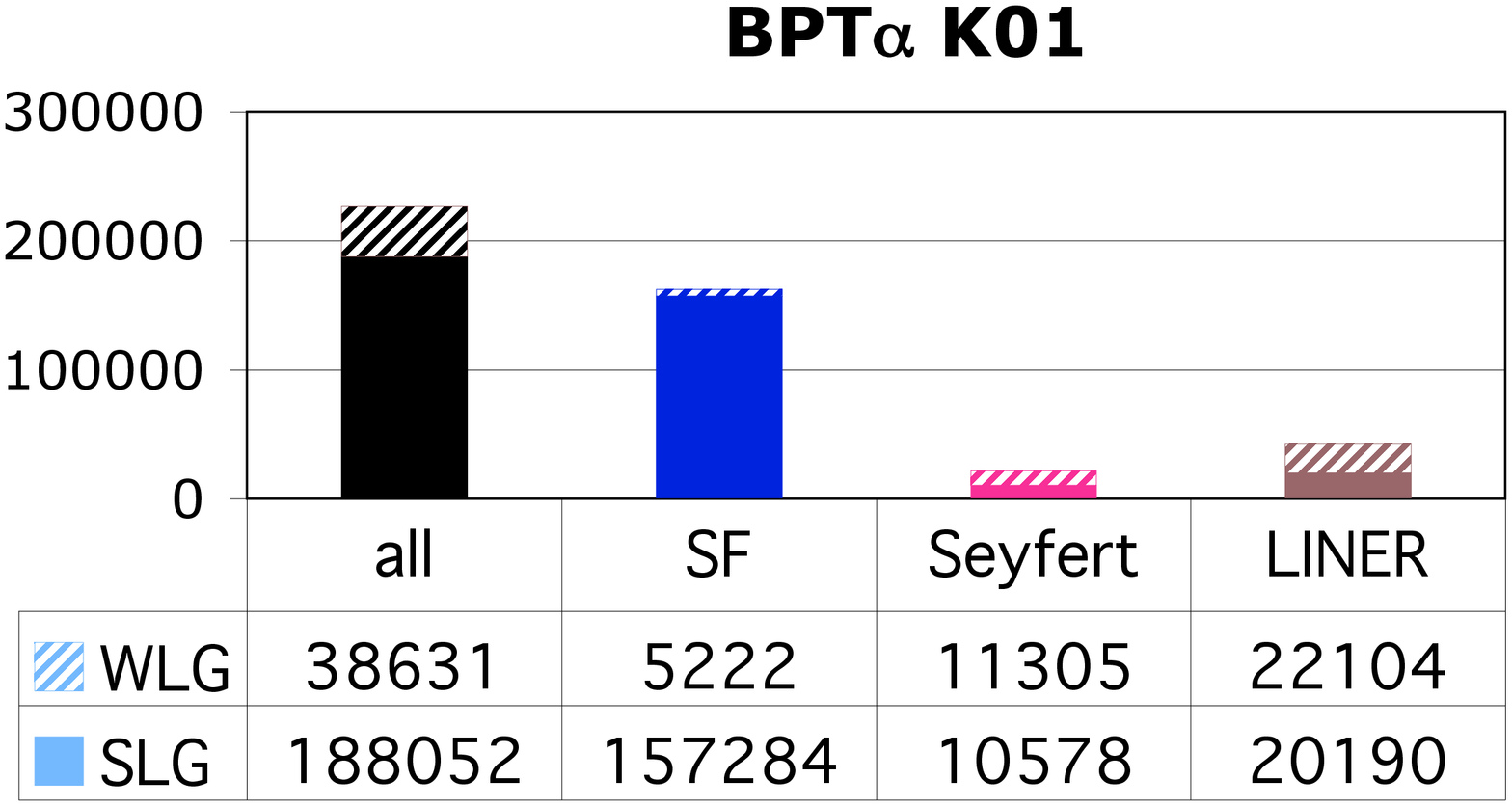}
    \includegraphics[width=.2\textwidth, angle=270, bb=100 50 530 710]{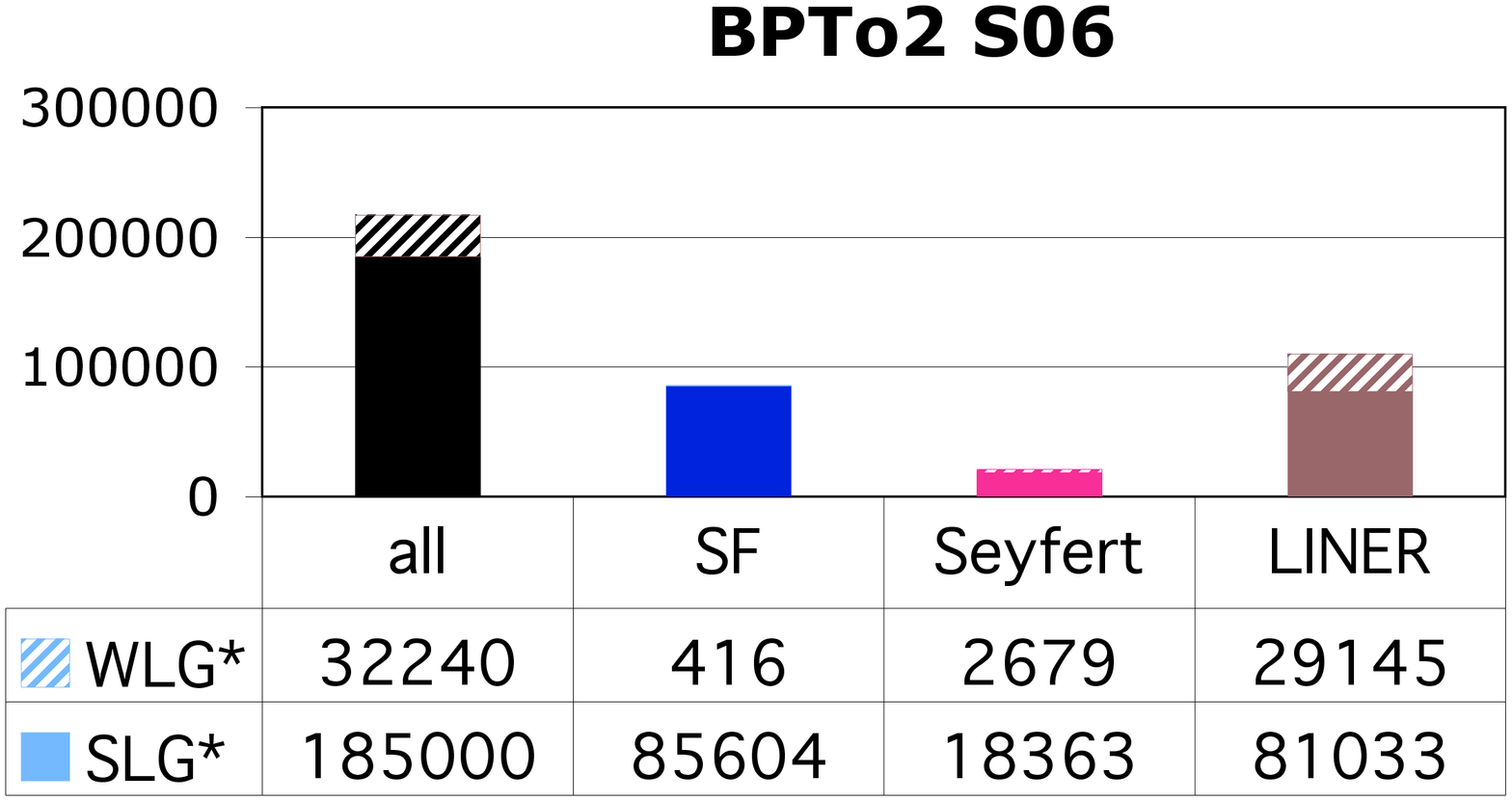}
    \includegraphics[width=.2\textwidth, angle=270, bb=100 50 530 710]{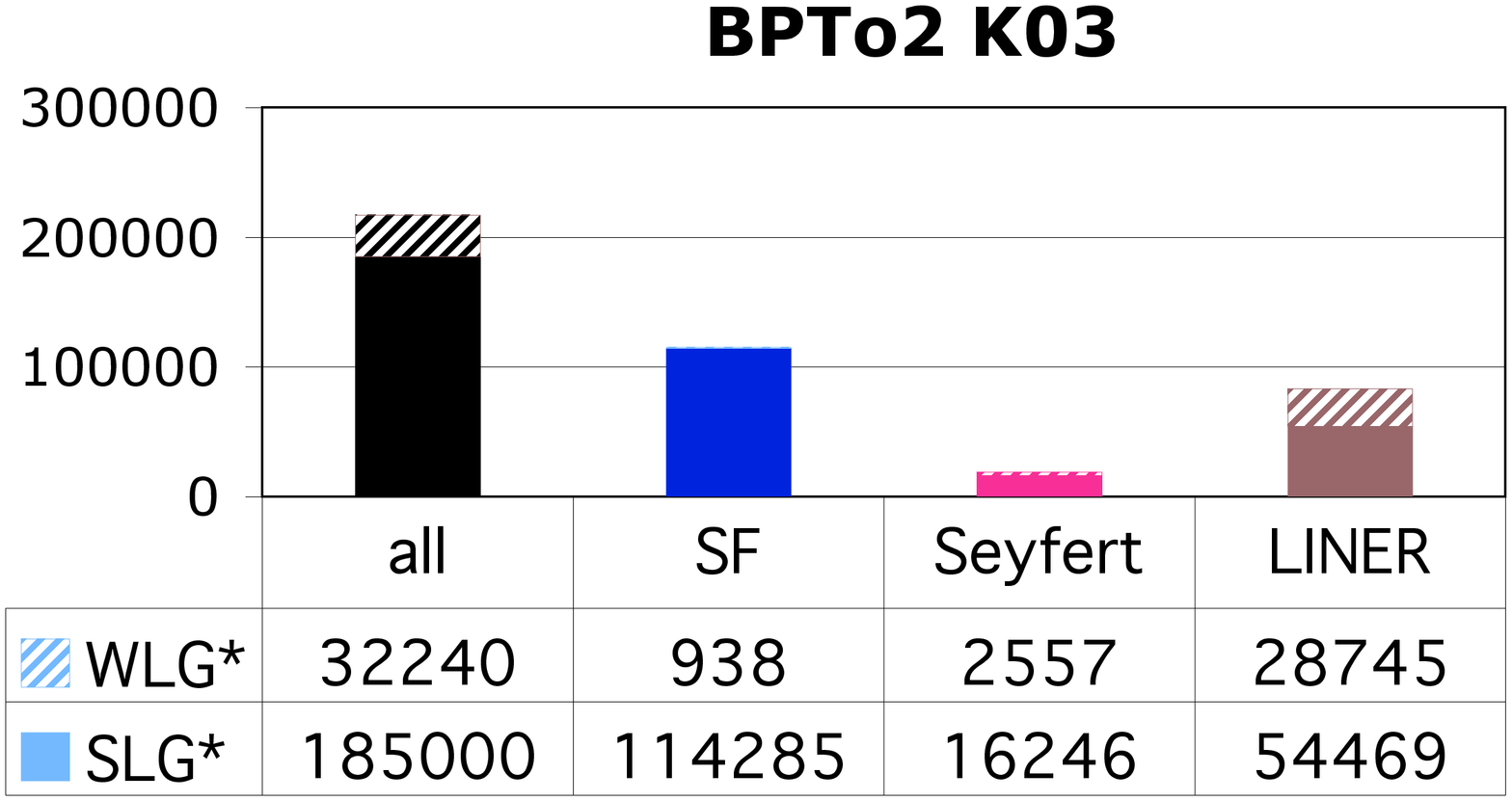}
    \includegraphics[width=.2\textwidth, angle=270, bb=100 50 530 710]{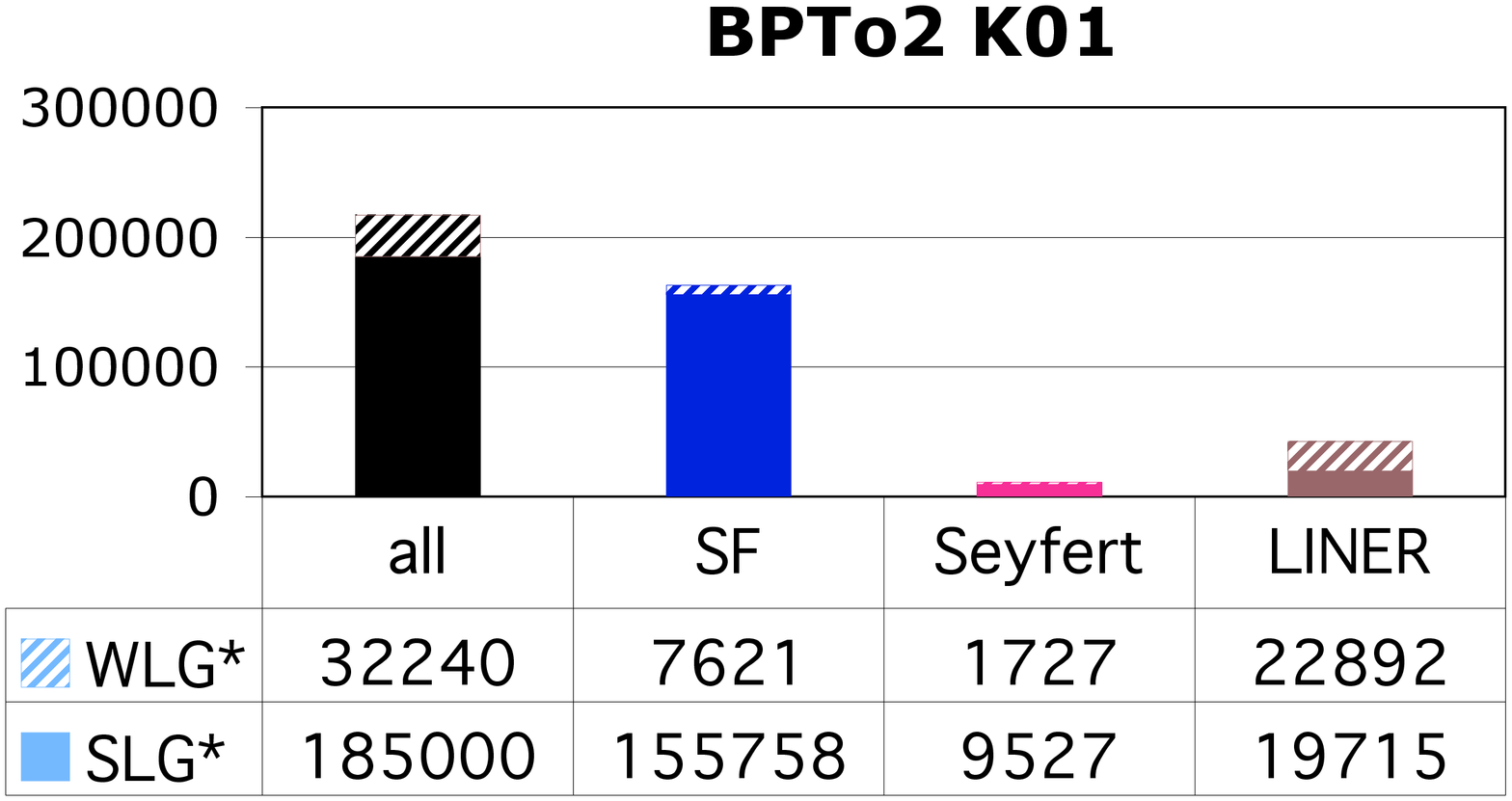}
    \includegraphics[width=.2\textwidth, angle=270, bb=100 50 530 710]{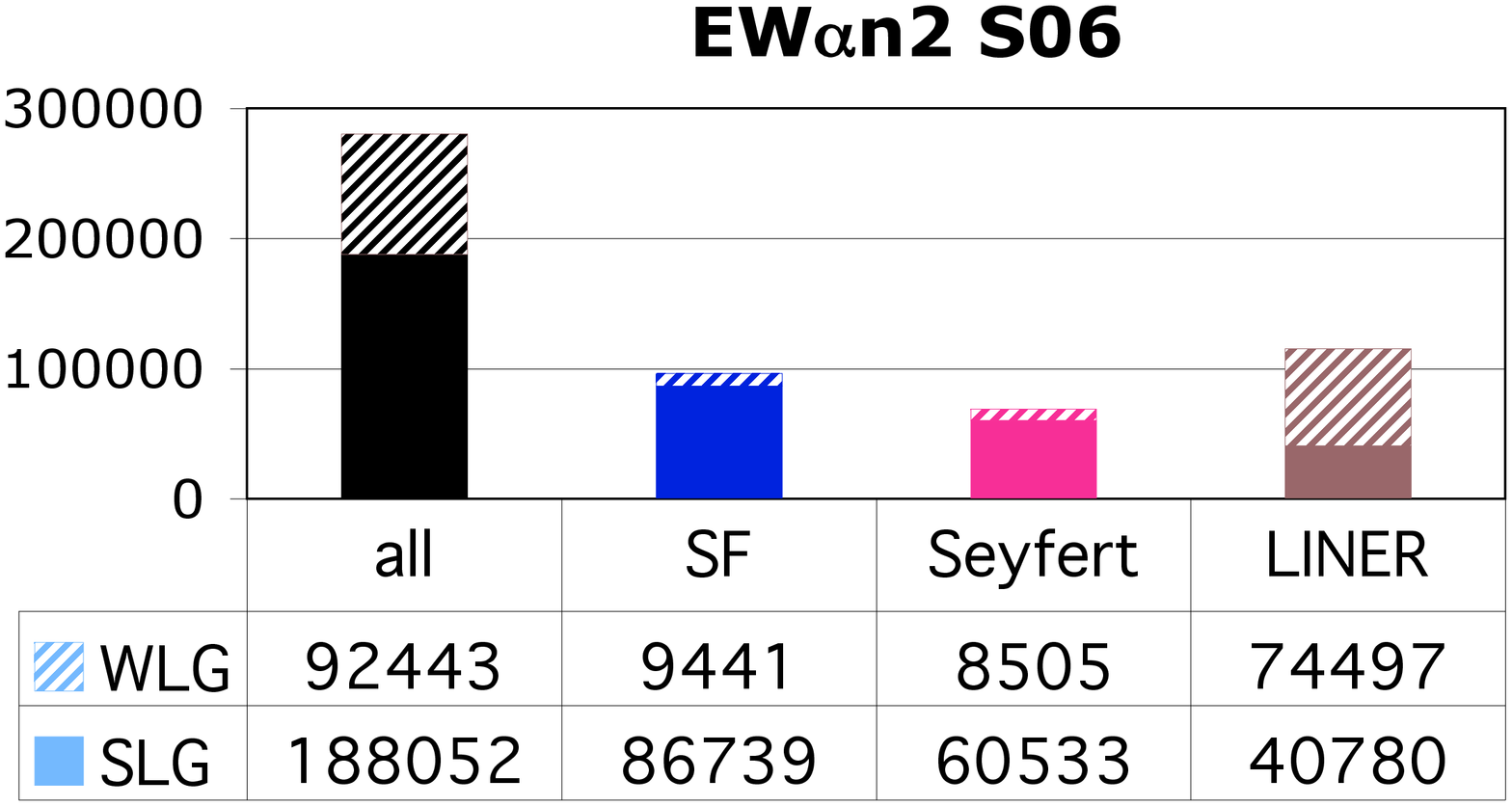}
    \includegraphics[width=.2\textwidth, angle=270, bb=100 50 530 710]{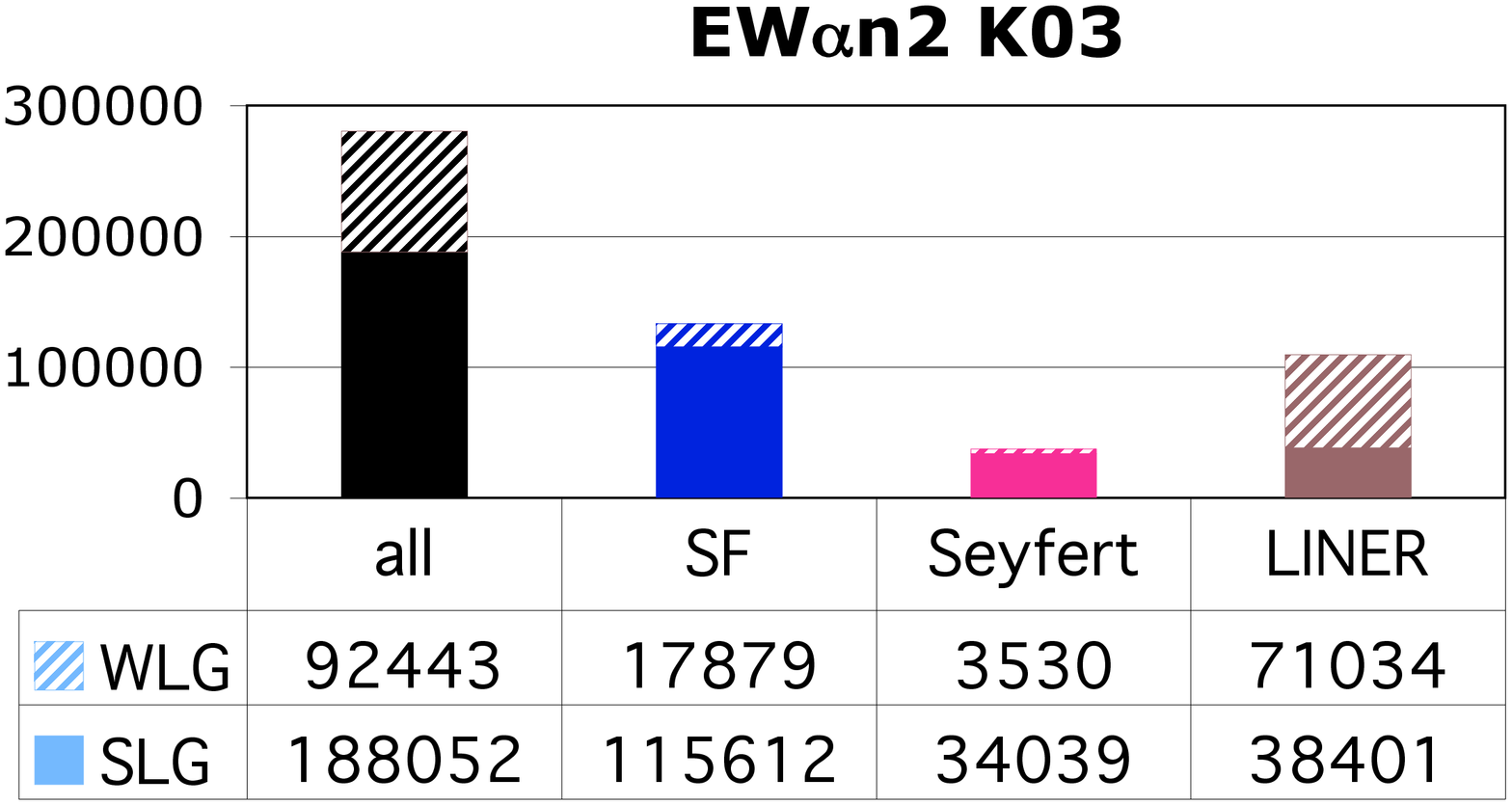}
    \includegraphics[width=.2\textwidth, angle=270, bb=100 50 530 710]{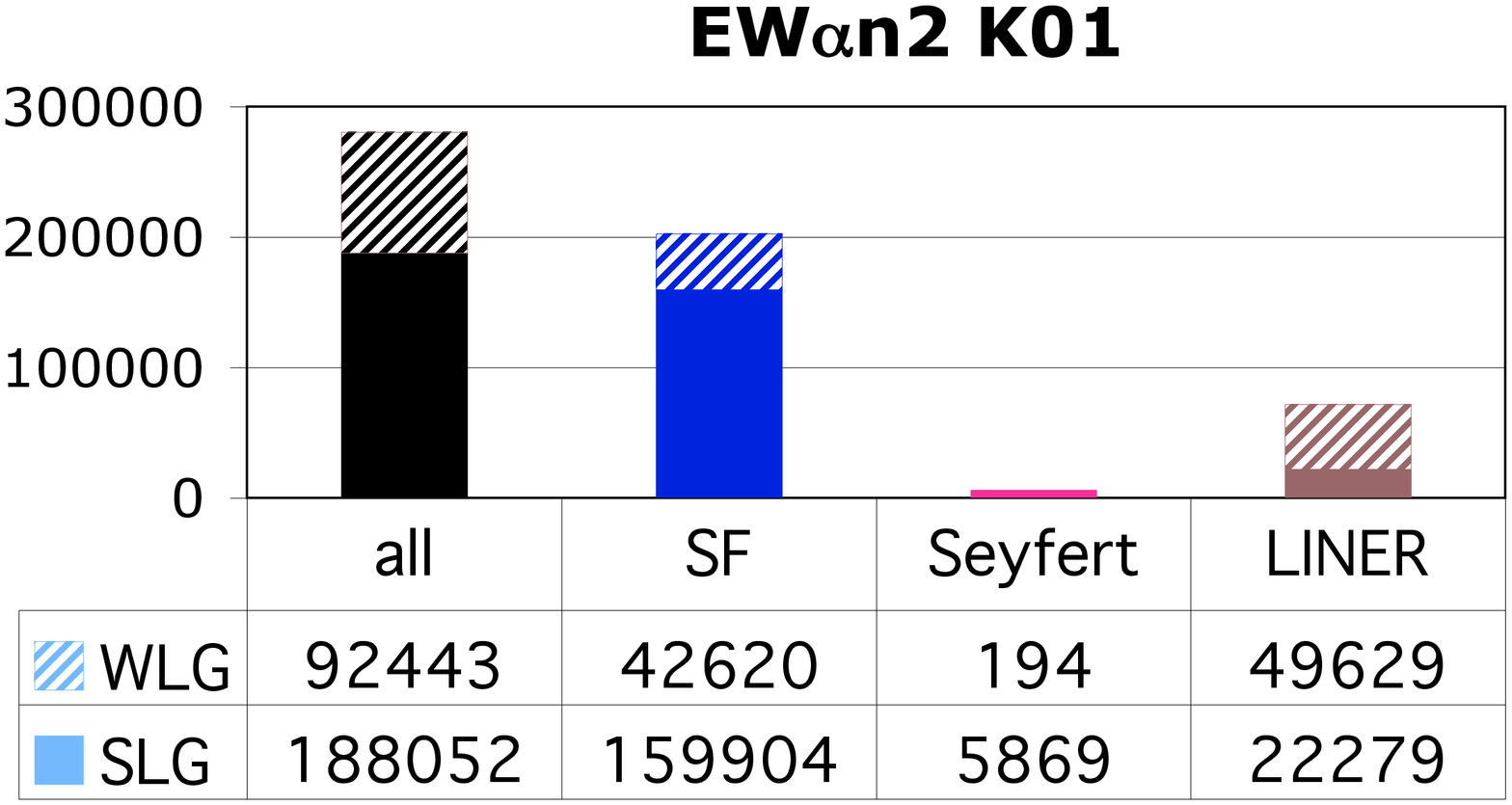}
\caption{As Fig.~\ref{fig:classif_wlgs}, but now including both SLGs
and WLGs to provide a global census of ELGs classes in the SDSS.}
\label{fig:census_alternative}
\end{figure*}

In order to have a reference for comparison, we show in Fig.\
\ref{fig:census_O3Hb} the result of the classification of our SLG
sample into SF, Seyfert and LINER-like classes using the BPT diagram
and the S06 (left), K03 (middle) and K01 (right) divisory lines. The
relative populations of SF/Seyferts/LINERs in our SLG sample are, in
percentages: 44/6/50 (S06 scheme), 60/6/34 (K03) and 84/5/11
(K01). Thus, LINERs represent half of the total population of SLGs
when using the S06 scheme, and about one third when using the K03
one. This large difference for such apparently equivalent
classification schemes stems from the large number of sources in the
geometrically small space between the S06 and K06 lines in the BPT
diagram. The K01 scheme gives a vast majority of SF galaxies, but
whereas in the S06 and K03 schemes the SF class is meant to represent
``pure SF'', K01-SFs include everything below their ``extreme
starburst'' line, whose meaning has been questioned above. Obviously,
these differences must be kept in mind when comparing numbers of
galaxies derived under different schemes.

Fig.~\ref{fig:census_alternative} shows the distribution among SF,
Seyferts and LINERs of all the galaxies whose emission line
intensities allow a decent ($SN_\lambda \ge 3$) classification in the
BPT$\alpha$, BPTo2 or EW$\alpha$n2 diagrams.  As in the previous
figures, S06, K03 and K01 classifications are shown separately.

By using the BPT$\alpha$ (top panels) instead of the BPT, we are able
to add 21\% more objects to our census. SLGs are plotted as filled
areas, while WL-Hs are represented by hatched areas. As expected,
these WLGs do not increase much the SF population (except in the K01
scheme), but they significantly increase the Seyfert and LINER
populations.  The BPTo2 diagram (middle panels in
Fig.~\ref{fig:census_alternative}) adds 17\% more galaxies with
respect to those classifiable in the BPT.  The partitioning is similar
to that obtained with the BPT$\alpha$ diagram, except for a somewhat
smaller proportion of Seyferts.  Whereas in the BPT$\alpha$ diagram
weak line Seyferts make up between 45 (S06) and 52\% (K01) of the
total WLG + SLG Seyfert population, in the BPTo2 this fraction is
between 13 and 15\%.  Given the higher Seyfert/LINER diagnostic power
of the BPTo2, we favor the latter results, which corroborate the view
that few WLGs have Seyfert-like emission lines.  In both diagrams the
K01 scheme is the one for which WLGs most increase the population of
LINERs (by 52\% in the BPT$\alpha$ and 54\% in the BPTo2). The
implication is that, under the widespread view that the K01 criteria
isolate ``pure AGN'', a full half of the ``pure-LINER'' population is
completely missed by imposing a standard $SN_\lambda \ge 3$ quality
control on the BPT lines.

As a whole, classifications derived with the BPT$\alpha$ and BPTo2
diagrams do not change drastically the balance of SF, Seyfert and
LINER galaxies. For instance, in the S06 scheme, the SF/Seyfert/LINER
percentages change from 44/6/50 with the BPT, to 37/13/50
(BPT$\alpha$) and 40/10/50 (BPTo2).

Larger differences are obtained classifying galaxies with the
EW$\alpha$n2 diagram, specially for the S06 and K01 schemes, which
yield 34/25/41 and 72/2/26 percentage SF/Seyfert/LINER proportions.
The large fraction of S06-Seyferts comes mostly from the $W_{\Ha} > 6$
\AA\ objects close to the $\log \nii/\Ha = -0.40$ frontier, whose \Ha
emission is likely contaminated (if not dominated) by star-formation
(see also Fig.\ \ref{fig:DDs_X_EWHaEQ6Cuts}).  In the K01-scheme, the
increase in the LINERs share is essentially due to WLGs, while the
miniscule fraction of Seyferts is due to the fact that the $\log
\nii/\Ha = -0.10$ frontier is a rather inefficient translation of the
K01 SF/AGN criteria. In the K03 scheme, on the other hand, the changes
in the SF/Seyfert/LINER classes deduced from the EW$\alpha$n2 diagram
are not as large. The inferred proportions are 48/13/39, compared to
60/6/34 deduced with the BPT.

A general result which is immune to the subtleties associated to all
these comparisons is that, in all diagrams and for all classification
schemes, LINERs are the ones which grow the most in absolute numbers
with the inclusion of WLGs. This is also the class which grows the
most in relative terms in the two diagrams most apt to distinguish AGN
subtypes: the BPTo2 and EW$\alpha$n2.

It seems to us that, for practical purposes, the EW$\alpha$n2
classification is more convenient, not only because it allows one to
increase by about 50\% the number of galaxies that can be classified,
but also because it considers the strength of the AGN with respect to
its host galaxy as a parameter that is more important than the
ionization state of the gas to distinguish between strong and weak
AGN.

\subsection{The Seyfert/LINER dichotomy revisited}
\label{sec:K06Bimodality}

\begin{figure*}
\includegraphics[bb= 50 200 570 700,width=\textwidth]{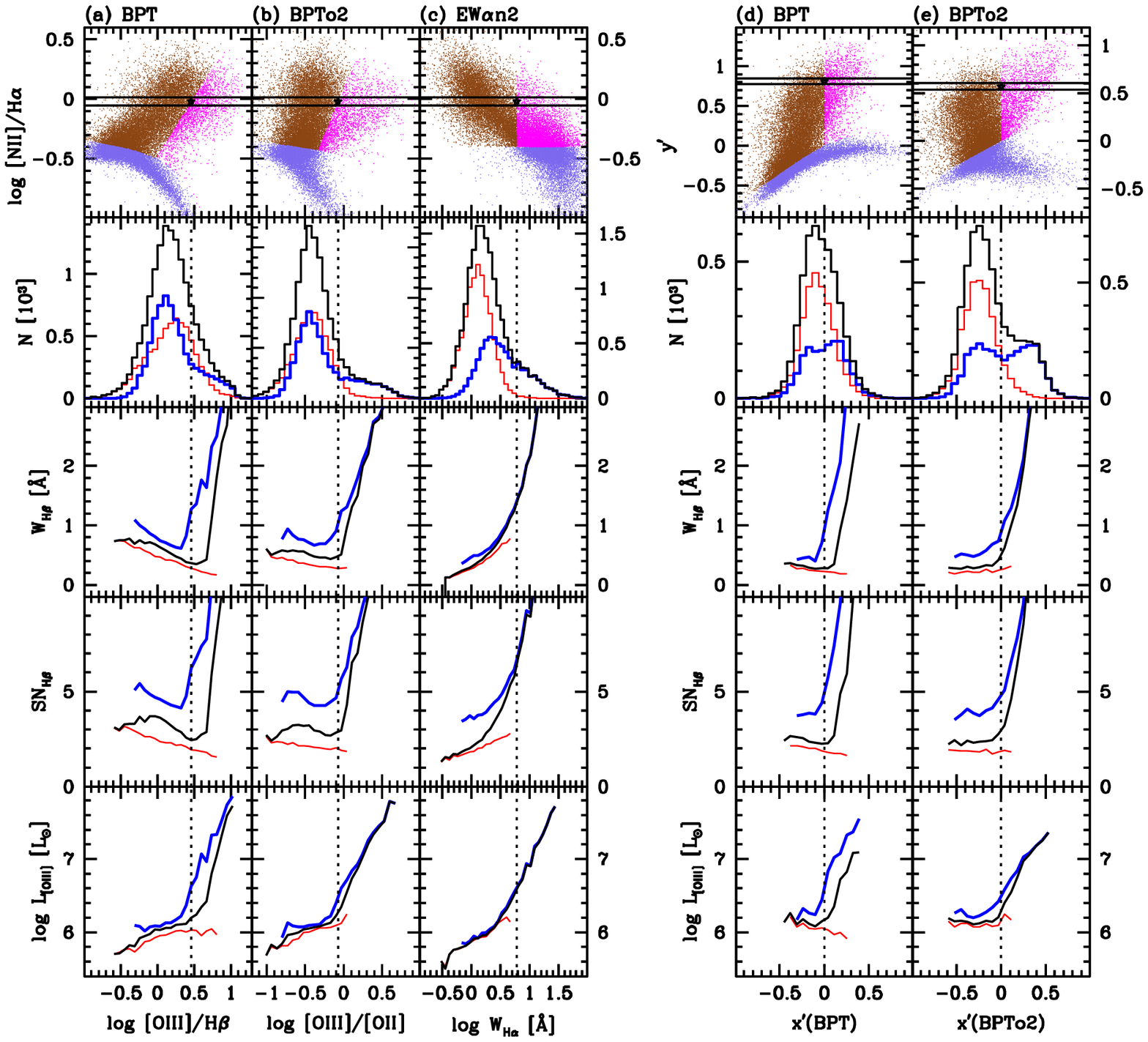}
\caption{A different look at the Seyfert/LINER bimodality in
diagnostic diagrams.  The top panels in rows (a), (b) and (c) show the
BPT, BPTo2 and EW$\alpha$n2 diagrams, but with $\log \nii/\Ha$ on the
$y$-axis. Violet, brown and magenta points correspond to SF, Seyferts
and LINERs as classified with the S06 scheme.  The second panel from
the top show histograms of $x$-values for SLGs (blue), WLGs (red) and
all ELGs (black) galaxies within the narrow strip in $\log \nii/\Ha$
marked in the top panel.  Dotted vertical line marks the $x$-value
associated to the center of the selected $y$-range on the
corresponding Seyfert/LINER divisory line.  In all cases this
coincides with a shoulder in the SLG histograms.  The bottom panels
show the median value of the equivalent width and signal-to-noise
ratio of \Hb, and the \oiii luminosity, separated in SLGs, WLGs and
all ELGs. Panels (d) and (e) are like panels (a) and (b) but applying
translations and rotations to the BPT and BPTo2 diagrams.  The
transformations from the original coordinates to the new ones
$(x^{\prime},y^{\prime})$ is such that {\em (i)} the origin is
centered at the intersection of the S06 SF/AGN and Seyfert/LINER
divisory lines, and {\em (ii)} the Seyfert/LINER line becomes vertical
($x^\prime = 0$). This rotation ensures that the narrow strips in
$y^\prime$ (marked in the top panels) cross the Seyfert/LINER line
perpendicularly, facilitating the visualization of the bimodalities.}
\label{fig:Bimod}
\end{figure*}

The Seyfert/LINER classification criteria proposed in this paper
simply transpose the bimodality identified by K06 to our more economic
diagnostic diagrams.  Given that the stringent requirements on data
quality imposed by K06 completely exclude the huge population of
AGN-like WLGs, it is fit to ask: Does the Seyfert/LINER dichotomy
subsist when WLGs are considered?

A visually convenient way to inspect bimodality effects on our
diagnostic diagrams is to swap the order of the axis and count sources
as a function of $x$ for a fixed $y = \log \nii/\Ha$.  This is done in
Figs.\ \ref{fig:Bimod}a--c for the BPT, BPTo2 and EW$\alpha$n2
diagrams. The top panels show the inverted diagrams, color coding
S06-SF, S06-Seyferts and S06-LINERs by violet, magenta and brown
points, respectively. Horizontal lines mark a 0.07 dex wide window
around $y = -0.02$, chosen for illustration purposes. Histograms of
SLGs (blue), WLGs (red) and all ELGs (black) along this narrow strip
are presented in the second panel from the top, while the bottom
panels show the corresponding median equivalent width and signal to
noise ratio of \Hb, and the \oiii luminosity as a function of $x$.

In the BPT diagram (Fig.\ \ref{fig:Bimod}a), the distribution of SLGs
presents a clear shoulder at $\log \oiii/\Hb = 0.46$, exactly the
value corresponding to the transition from LINERs to Seyferts for
$\log \nii/\Ha = -0.02$ in equation \ref{eq:TranspDivLine_K06_BPT}
(marked by a vertical dotted line). This evidence for a bimodality
{\em disappears} when WLGs are included, as shown by the black
histogram.  The $W_{\Hb}$ and $SN_{\Hb}$ panels shown that the
Seyfert/LINER frontier coincides with the region where \Hb is the
weakest, and thus also where restrictions upon $SN_{\Hb}$ have a
larger impact.  Does this mean that there is a continuum of emission
line properties between Seyferts and LINERs when considering a more
complete sample of galaxies than SLGs, and that there is no dichotomy?

It does not seem to be the case. In the BPTo2 diagram (Fig.\
\ref{fig:Bimod}b), the Seyfert/LINER bimodality survives the inclusion
of WLGs. Again, SLGs have a shoulder in the distribution of $x$-values
starting at the value of \oiii/\oii expected from our transposed
version of the K06 Seyfert/LINER classification (equation
\ref{eq:TranspDivLine_K06_BPTo2}).  As in the BPT, WLGs approximately
double the low $x$ ($= \log \oiii/\oii$ in this case) counts, but this
time they do {\em not} wash away the change in the distribution as one
goes from LINERs to Seyferts. The histograms for the EW$\alpha$n2
diagram (Fig.\ \ref{fig:Bimod}c), also show evidence for two
populations, with a high $W_{\Ha}$ hump starting precisely at our
proposed the Seyfert/LINER frontier: $W_{\Ha} = 6$ \AA.

Figs.\ \ref{fig:Bimod}d and e show two further experiments with the
BPT and BPTo2 diagrams.  The top panels show these diagrams after
translation and rotation operations. The transformations from the
original coordinates to the new ones $(x^{\prime},y^{\prime})$ are
such that {\em (i)} the origin is centered at the intersection of the
SF/AGN and Seyfert/LINER divisory lines, and {\em (ii)} the
Seyfert/LINER line becomes vertical ($x^\prime = 0$). This allows us
to count galaxies across a $y^\prime$-strip which crosses our
Seyfert/LINER frontiers at right angles, facilitating the
visualization of bimodalities. Indeed, unlike with the original
coordinates, the histograms for SLGs now show two modes, particularly
strong in the BPTo2. These alternative representations of the data
confirm that the bimodality in the BPT appears only for SLGs, whereas
in the BPTo2 it cannot be attributed to a selection effect.
Interestingly, the $W_{\Hb}$, $SN_{\Hb}$ and $L_{\oiii}$ profiles
become essentially flat for $x^{\prime} < 0$ (corresponding to
LINERs), signalling a transition to a different regime.

These results suggest that there are, indeed, at least two classes of
AGN-like galaxies.  K06 speculate that this dual behaviour is
analogous to the high and low states of black hole accretion in X-ray
binaries, ie., two regimes of a same physical process.  A different
possibility is that the bimodality in emission line properties
originates from two completely different sources of ionizing
radiation: an active nucleus versus old stars.  As shown by
\citet{Stasinska_etal_2008}, many SDSS galaxies belonging to the LINER
zone in the BPT diagram have an old stellar population whose ionizing
photons alone are able to produce the observed emission line
intensities. These ``retired galaxies'' are erroneously counted as
AGN, leading to the illusion of a dichotomy in the AGN population. It
might therefore be that the real dichotomy is between AGN and retired
galaxies, and not between two states of black hole accretion. A more
detailed assessment of this scenario is postponed to a forthcoming
communication.


\section{Summary and conclusion}
\label{sec:Conclusions}

This paper revisited the emission line classification of galaxies,
focusing on the numerous population of galaxies which are often left
out of emission line studies because of their weak (low
signal-to-noise) lines. We have shown that WLGs, defined as systems in
which both \Ha and \nii have $\ge 3$ sigma detections but where either
or both of \Hb and \oiii is weaker, amount to about 1 in 3 emission
line galaxies in the SDSS DR7. This already large fraction increases
nearly two-fold if one concentrates on the right wing of the \oiii/\Hb
versus \nii/\Ha BPT diagram, where ionizing sources other than young
stars have a significant impact on integrated emission line
properties. The lack of good quality data leaves these objects in a
classification limbo, preventing a complete census the population of
galaxies with emission lines.

In order to rescue WLGs from their uncertain place in current emission
line classification schemes, we have investigated alternative
diagnostic diagrams, all of which keep \nii/\Ha as a horizontal axis,
but where \Hb is replaced by a stronger line (\Ha or \oii), or where
the ionization-level sensitive \oiii/\Hb ratio is replaced by the
equivalent width of \Ha. The classification power of these ``cheaper''
alternative diagrams was evaluated from the location of well
classified, strong line sources in these same diagrams.

To avoid introducing further entropy in the emission line taxonomy of
galaxies, SF/AGN and Seyfert/LINER border lines in these diagrams were
traced by means of an objective method devised to transpose popular
classification schemes (K01, K03, S06, and K06) in an optimal way.  We
find that the BPT$\alpha$ (Fig.\ \ref{fig:WLGsOnBPTa}) and BPTo2
(Fig.\ \ref{fig:WLGsOnBPTo2}) diagrams do an excellent job in placing
galaxies with weak \Hb but strong \oiii in standard SF/AGN and
Seyfert/LINER categories.  In particular, the BPTo2 has the ability of
mimicking with exquisitely high efficiency the detailed Seyfert/LINER
classification of K06, but at a much reduced cost in terms of data
quality requirements.

The most economic classification is achieved with the EW$\alpha$n2
diagram ($W_{\Ha}$ versus \nii/\Ha), where SF and AGN are segregated
in terms of \nii/\Ha, while Seyferts and LINERs differ in the vertical
axis, with an optimal class division at $W_{\Ha} = 6$ \AA. Inevitably,
the correspondence with standard classification schemes is not as good
as for diagrams involving more data, but the cost/benefit is such that
it is tempting to propose it as a fundamental classification
scheme. This is the only scheme able to classify all galaxies with 3
sigma or better detection of \Ha and \nii, which comprise $\sim 3/4$
of the galaxies is the SDSS.

Working definitions were proposed to identify which of \Hb and/or
\oiii prevents a full BPT-based classification, leading to the weak
\Hb but strong \oiii (WL-H), weak \oiii but strong \Hb (WL-O), and
weak \Hb and \oiii (WL-HO) sub-divisions. WL-H and WL-HO sources share
many emission line properties which indicate their predominantly
LINER-like nature.  WL-Os, on the other hand, are best described as SF
systems whose \oiii flux is suppressed by enhanced nebular cooling
associated with high metallicity. Some of them, however, mingle
amongst WL-H and WL-HOs, and thus probably have an AGN-like component.

Our WLG-rescuing operation leads to a revision of the partition of
galaxies in the nearby Universe into different spectral types in the
sense that metal-rich SF galaxies and especially LINERs occur more
often than would deduced ignoring WLGs.

While adapting currently popular schemes to distinguish SF galaxies
from AGN we have stumbled upon known, but widely overlooked,
inconsistencies. The most worrying of these is the way that the K01
``extreme starburst'' line in the BPT diagram is currently used to
isolate ``pure AGN''. This line was never designed to do this, neither
is it capable of doing so in more than a qualitative sense.
Similarly, and contrary to current practice, using the K01 line as an
upper bound in conjunction with any plausible SF/AGN lower boundary in
the BPT plane is {\em not} a safe method to identify systems where
both star-formation and an AGN contribute significantly to the
emission lines (i.e., ``SF+AGN composites'').  Overall, we feel it is
preferable to abandon the K01 line altogether instead of paying the
price of erroneously interpreting the physical nature of the dominant
ionizing source in galaxies.

We have also revisited the Seyferts vs LINERs dichotomy in emission
line properties. In the BPT diagram, the inclusion of the
``forgotten'' WLGs practically erases this duality, but more robust
diagrams like the BPTo2 and the EW$\alpha$n2 support the existence of
two populations of AGN-like galaxies. We speculate that this dichotomy
is not inherent to AGN themselves, but a consequence of mistaking
retired galaxies for AGN. The ionizing source in retired galaxies is
stellar, yet these systems look like AGN in all properties accessible
with SDSS data.  Regardless of whether this interpretation of the
origin of the Seyfert/LINER dichotomy is correct, retired galaxies are
a natural consequence of stellar evolution and are surely counted as
AGN in the SDSS and similar surveys. Further work is needed to
identify these fake AGN.


\section*{ACKNOWLEDGMENTS}

The {\sc starlight} project is supported by the Brazilian agencies CNPq,
CAPES, FAPESP, by the France-Brazil CAPES-COFECUB program, and by
Observatoire de Paris. 

Funding for the SDSS and SDSS-II has been provided by the Alfred
P. Sloan Foundation, the Participating Institutions, the National
Science Foundation, the U.S. Department of Energy, the National
Aeronautics and Space Administration, the Japanese Monbukagakusho, the
Max Planck Society, and the Higher Education Funding Council for
England. The SDSS Web Site is http://www.sdss.org/.

The SDSS is managed by the Astrophysical Research Consortium for the
Participating Institutions. The Participating Institutions are the
American Museum of Natural History, Astrophysical Institute Potsdam,
University of Basel, University of Cambridge, Case Western Reserve
University, University of Chicago, Drexel University, Fermilab, the
Institute for Advanced Study, the Japan Participation Group, Johns
Hopkins University, the Joint Institute for Nuclear Astrophysics, the
Kavli Institute for Particle Astrophysics and Cosmology, the Korean
Scientist Group, the Chinese Academy of Sciences (LAMOST), Los Alamos
National Laboratory, the Max-Planck-Institute for Astronomy (MPIA),
the Max-Planck-Institute for Astrophysics (MPA), New Mexico State
University, Ohio State University, University of Pittsburgh,
University of Portsmouth, Princeton University, the United States
Naval Observatory, and the University of Washington.


\bibliographystyle{mn2e}

\end{document}